\documentclass[aps,prb,twocolumn,superscriptaddress,longbibliography]{revtex4-2}

\usepackage{amsmath,amssymb,amsfonts}
\usepackage{bm}
\usepackage{graphicx}
\usepackage{subcaption}

\newcommand{\dd}{\mathrm{d}}
\newcommand{\ii}{\mathrm{i}}
\newcommand{\cl}{\mathrm{cl}}
\newcommand{\q}{\mathrm{q}}

\begin{document}

\title{Hydrodynamic Cooperons in Electron Fluids:
Schwinger--Keldysh Derivation and Quantum Corrections to Magnetoresistance}

\author{Alberto Cortijo}
\email{alberto.cortijo@csic.es}
\affiliation{Instituto de Ciencia de Materiales de Madrid (ICMM), Consejo Superior de Investigaciones Científicas (CSIC), Sor Juana Inés de la Cruz 3, 28049 Madrid, Spain}


\begin{abstract}
We develop a Schwinger--Keldysh effective theory for quantum-interference corrections in a two-dimensional electron system in the hydrodynamic regime. Starting from the clean hydrodynamic fixed point, we introduce a minimal random-friction disorder model that generates a finite momentum-relaxation time within the self-consistent Born approximation. The disorder-averaged theory then allows us to construct a hydrodynamic Cooperon and to compute the associated self-energy corrections to the collective modes. Conservation laws protect the density and momentum sectors, so that the leading quantum-coherence correction is forced into the spin-two stress sector. The associated stress self-energy renormalizes the shear viscosity and modifies both the Gurzhi response and its low-field magnetohydrodynamic signatures.
\end{abstract}

\maketitle


\section{Introduction}

Quantum interference in disordered conductors is usually formulated around the diffusive Cooperon, the retarded--advanced ladder generated by pairs of time-reversed electronic paths.  In two dimensions this infrared mode produces the weak-localization correction to charge transport and is strongly cut off by dephasing effects and magnetic fields \cite{Lee85,Altshuler82,Hershfield86,Bergmann84}.  Electron hydrodynamics is organized by a different infrared fixed point.  When momentum-conserving collisions dominate over impurities, phonons, and boundary relaxation events, the long-lived variables are collective charge and momentum densities, and transport is controlled by viscous flow rather than by single-particle diffusion.  This hydrodynamic regime underlies the Gurzhi effect, viscous backflow, nonlocal response, and hydrodynamic magnetotransport observed or analyzed in high-mobility two-dimensional conductors \cite{Gurzhi63,deJong95,Molenkamp94,Gurzhi95,Bandurin16,KrishnaKumar17,Berdyugin19,Lucas18,Narozhny17,Fritz24,Varnavides23}.  The problem addressed in this paper is how a Cooperon-like interference channel can be consistently embedded in such a hydrodynamic theory.

This hydrodynamic fixed point prevents a direct transposition of the conventional weak-localization correction to the conserved hydrodynamic modes. Charge and momentum conservation impose Ward identities that forbid a uniform dissipative self-energy for the scalar and vector sectors. Any admissible coherence correction must therefore enter a nonconserved hydrodynamic sector, with the spin-two stress mode providing the leading possibility \cite{Lucas18,Narozhny17,CrossleyGloriosoLiu17,GloriosoLiu18,Fritz24}.  In the angular-harmonic notation used in this work, the protected modes are the density harmonic $m=0$ and the momentum harmonics $m=\pm1$.  A quantum-interference correction acting directly as a mass in either sector would violate the conservation-law structure of the clean fixed point.  The first admissible target is therefore the leading nonconserved rotational channel, the spin-two $m=\pm2$ stress sector.  The hydrodynamic Cooperon constructed here renormalizes the stress relaxation rate and hence the viscosity, rather than the charge conductivity directly.  This is the central difference between the present mechanism and ordinary diffusive weak localization.

The first purpose of this work is to provide the technical construction behind this statement.  We formulate the clean hydrodynamic harmonic hierarchy as a real-time Schwinger--Keldysh (SK) effective theory, because the SK language keeps retarded response, Keldysh fluctuations, and the fluctuation--dissipation relation in a single causal framework \cite{Kamenev23,CrossleyGloriosoLiu17,GloriosoLiu18}.  Static disorder is introduced as a random-friction field coupled locally to the momentum density.  This is not intended as a universal microscopic disorder model; it is a minimal hydrodynamic vertex that weakly relaxes momentum while preserving the clean scalar, vector, and spin-two mode structure.  After Gaussian disorder averaging, the same vertex produces a self-consistent Born momentum-relaxation rate and a maximally crossed retarded--advanced ladder of collective shear propagators, in analogy with the Cooperon ladder of disordered metals \cite{Lee85,Vollhardt80b,Altshuler82,Hershfield86}.  The internal lines in this diagrammatic way of thinking, however, are hydrodynamic momentum modes rather than microscopic fermions.

The second purpose is to extend the construction to the presence of a weak perpendicular magnetic field, which is the main new ingredient of the present paper. The magnetic field first changes the clean hydrodynamic fixed point: the vector harmonics precess at the cyclotron frequency, the spin-two harmonics precess at twice that frequency, and eliminating the stress sector produces both longitudinal and Hall viscosities \cite{Alekseev16,Alekseev20,Avron98,Scaffidi17,Pellegrino17,Delacretaz17,Berdyugin19}.  Only after this magnetohydrodynamic fixed point is identified can one construct the magnetic hydrodynamic Cooperon.  The classical Lorentz shift modifies the hydrodynamic propagators and the Cooperon pole coefficients, while the genuine orbital Cooperon mass follows from gauge covariance of the center-of-mass coordinate, as in the magnetic cutoff of conventional weak localization \cite{Lee85,Bergmann84,Altshuler82,Hershfield86}.  The resulting field-dependent stress correction determines the renormalized longitudinal and Hall viscosities and the channel magnetoresistance.

This structure gives the magnetic field a double role.  At the semiclassical level, it rotates momentum and stress and thereby generates the usual magnetoviscous tensor.  At the interference level, it suppresses the hydrodynamic Cooperon through orbital dephasing and tends to restore the bare viscosity.  The competition between these two effects produces low-field structures in magnetoviscous response and in the Gurzhi magnetoresistance.  Within this minimal hydrodynamic disorder theory, the robust prediction is the emergence of coherence-sensitive low-field structures tied to the stress-sector renormalization. Disorder strength, dephasing, cutoff matching, and boundary geometry determine the detailed amplitudes and therefore connect the theory to specific devices.

The paper is organized as follows.  Section~\ref{sec:section_2} introduces the hydrodynamic harmonic notation and the bare viscosity.  Sec.~\ref{sec:section_3} develops the SK formulation, the random-friction disorder and the resulting effective interaction.  Sec.~\ref{sec:section_4} derives the transverse SCBA momentum-relaxation rate.  Sec.~\ref{sec:section_5} constructs the hydrodynamic Cooperon and its infrared pole.  Sec.~\ref{sec:section_6} computes the stress-sector self-energy and the zero-field viscosity correction.  Sections~\ref{sec:section_7}, \ref{sec:section_8}, and \ref{sec:section_9} extend the fixed point, SCBA, Cooperon, and stress correction to nonzero magnetic field, including the orbital mass.  Sec.~\ref{sec:section_10} discusses the resulting experimental signatures in magnetoviscosity and channel magnetoresistance.  Sec.~\ref{sec:section_11} summarizes the physical picture.

Throughout the paper we use units $k_B=\hbar=1$ for thermal and hydrodynamic relaxation rates, so that $T$ is measured as a frequency.  We restore $\hbar$ explicitly in the orbital Cooperon mass, where gauge covariance fixes the charge-$2e$ Peierls substitution.


\section{Hydrodynamic harmonics and bare viscosity}
\label{sec:section_2}

\begin{figure}[t]
    \centering
    \includegraphics[width=\columnwidth]{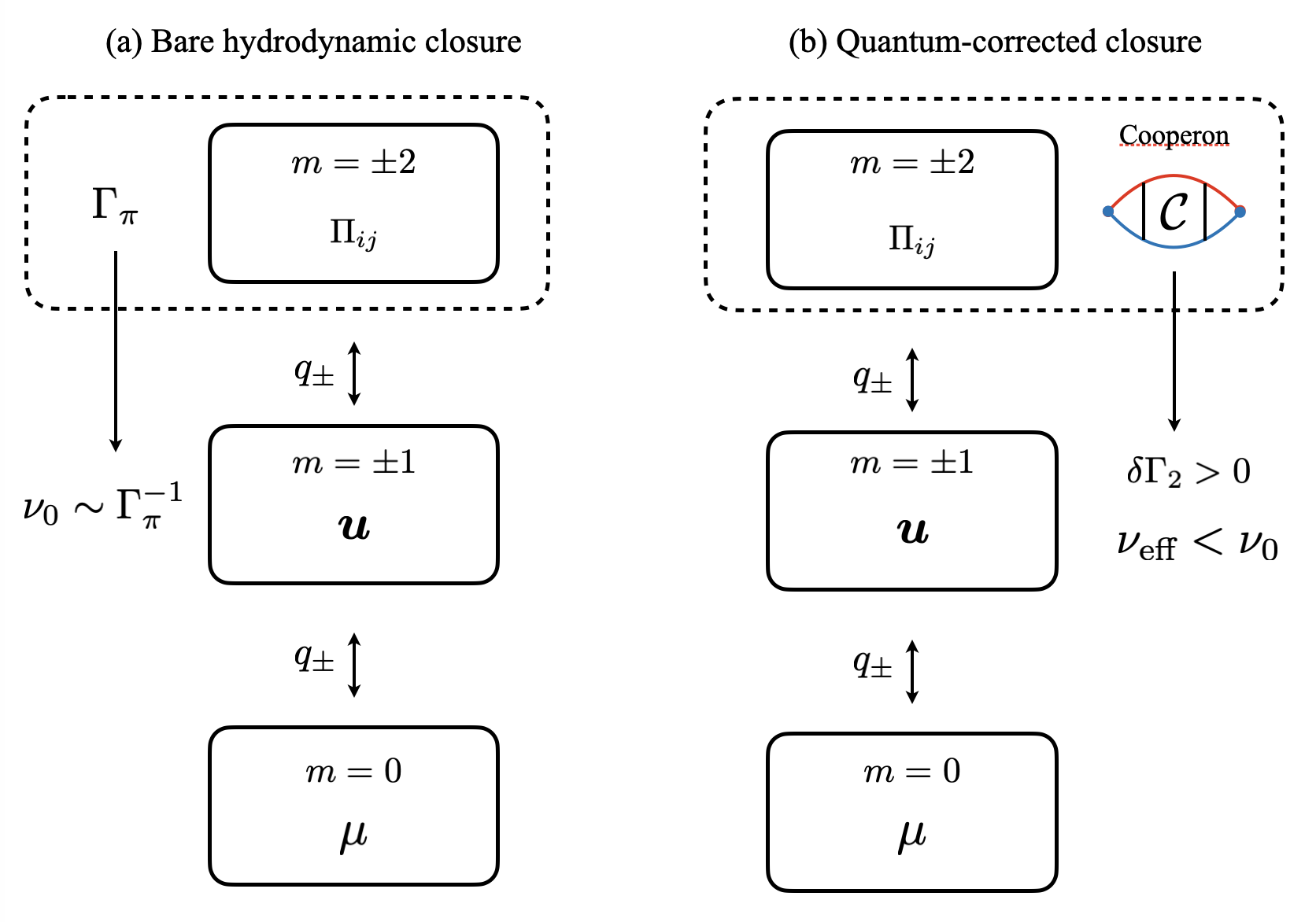}
    \caption{
    Hydrodynamic harmonic hierarchy and stress-sector closure.
    The density mode $m=0$ and momentum modes $m=\pm1$ are coupled by gradients
    and are protected by conservation laws, while the first nonconserved sector (represented by the dashed boxes) is
    the spin-two stress mode $m=\pm2$. In the bare hydrodynamic closure,
    integrating out the stress sector generates the viscosity $\nu_0$.
    In the quantum-corrected theory, the hydrodynamic Cooperon renormalizes the
    stress relaxation rate, $\Gamma_\pi \to \Gamma_\pi + \delta\Gamma_2$, with $\Gamma_\pi$ the bare spin-two relaxation rate, which feeds back
    into the momentum sector as an effective viscosity $\nu_{\mathrm{eff}}$.
    }
    \label{fig:mode_projection}
\end{figure}

Figure~\ref{fig:mode_projection} summarizes the central projection used throughout the paper.  The clean hydrodynamic hierarchy closes by eliminating the spin-two stress sector, while the quantum-corrected theory first renormalizes the stress relaxation rate and only then feeds back into the effective viscosity.  The figure also fixes the logic of the later sections: the conserved $m=0$ density and $m=\pm1$ momentum modes remain protected, whereas the nonconserved $m=\pm2$ stress modes provide the first channel in which a uniform Cooperon self-energy can appear.  The observable viscosity correction is therefore not inserted by hand at the Navier--Stokes level, but obtained by integrating out a stress sector whose relaxation rate has itself been dressed by the hydrodynamic Cooperon.

We first fix the zero-field hydrodynamic notation.  The kinetic language is useful because it makes the angular harmonics explicit, but the structure we need is more general: a scalar sector, a vector momentum sector, and a spin-two stress sector.  Eliminating the first nonconserved stress harmonics gives the bare shear viscosity used throughout the later Cooperon construction.

\subsection{Kinetic harmonics and bare hydrodynamic viscosity}

For an isotropic two-dimensional Fermi surface (parametrized by the angle $\theta$) with Fermi velocity $v_F$, the out of equilibrium distribution function is expanded as
\begin{eqnarray}
f(t,\bm x,\theta) = \sum_{m\in\mathbb Z} a_m(t,\bm x)e^{\ii m\theta}-f_{\mathrm{eq}}(\epsilon),
\end{eqnarray}
with the reality condition $a_{-m}=a_m^*$ and $f_{\mathrm{eq}}(\epsilon)$ denotes the distribution function in equilibrium.  The low angular harmonics retained below have the standard hydrodynamic interpretation: $m=0$ is the scalar density fluctuation (or chemical potential $\mu$), $m=\pm1$ is the momentum or velocity $\bm u$, and $m=\pm2$ are the spin-two stress harmonics.  Going from momentum $\bm P$ to velocity $\bm u$ will be done by using the momentum susceptibility $\chi_P$, $\bm P=\chi_P \bm u$. Higher harmonics are assumed to relax faster than the $m=\{0,\pm1,\pm2\}$ subspace and are integrated out at the order considered here.

The linearized kinetic equation,
\begin{equation}
\partial_t f+v_F\hat{\bm k}\cdot\nabla f=I_{\mathrm{coll}}[f],
\end{equation}
projects onto the harmonic hierarchy
\begin{equation}
\partial_t a_m+\frac{v_F}{2}\left(\partial_-a_{m-1}+\partial_+a_{m+1}\right)
=-\gamma_m^{(0)}a_m,
\label{eq:zero_field_hierarchy}
\end{equation}
where ($\partial_\pm \equiv \partial_x\pm\ii\partial_y$),
\begin{eqnarray}
\hat{\bm k}\cdot\nabla
&=&\frac{1}{2}\left(e^{\ii\theta}\partial_-+e^{-\ii\theta}\partial_+\right),
\\
\gamma^{(0)}_m a_m&=&-\int d\theta e^{-\ii m\theta}
I_{\mathrm{coll}}[a_m].
\end{eqnarray}
The clean fixed point conserves charge and momentum, so
\begin{equation}
\gamma_0^{(0)}=\gamma_{\pm1}^{(0)}=0,
\quad
\gamma_{\pm2}^{(0)}\equiv\tau_{\mathrm{mc}}^{-1}=\Gamma_{\pi}.
\end{equation}
Thus the spin-two sector is the first intrinsically relaxing sector when momentum conserving interactions dominate over momentum relaxing effects: no bare momentum-relaxing rate is inserted in the $m=\pm1$ equations in the hydrodynamic limit; momentum relaxation will be generated later by disorder averaging.

We now truncate Eq.~\eqref{eq:zero_field_hierarchy} to $m=0,\pm1,\pm2$, setting $a_{\pm3}=0$.  The truncation is controlled for
\begin{equation}
q\ell_{\mathrm{mc}}\ll1,
\quad
\omega\tau_{\mathrm{mc}}\ll1,
\quad
\ell_{\mathrm{mc}}=v_F\tau_{\mathrm{mc}}.
\end{equation}
In the notation of the companion Letter, this kinetic realization corresponds to
\begin{equation}
\Gamma_\pi\equiv\tau_{\mathrm{mc}}^{-1},
\quad
\ell_h=\ell_{\mathrm{mc}}=v_F\tau_{\mathrm{mc}},
\end{equation}
with the Fourier components $a_m(\omega,\bm q)$ defined by
\begin{equation}
a_m(t,\bm x)=\int\frac{\dd\omega}{2\pi}\frac{\dd^2q}{(2\pi)^2}
 e^{-\ii\omega t+\ii\bm q\cdot\bm x}a_m(\omega,\bm q),
\end{equation}
where $q_\pm=q_x\pm\ii q_y$.  The independent $m\ge0$ equations are
\begin{eqnarray}
\label{eq:mode_equations}
-\ii\omega a_0+\frac{\ii v_F}{2}\left(q_-a_{-1}+q_+a_1\right)&=&0,\\
-\ii\omega a_1+\frac{\ii v_F}{2}\left(q_-a_0+q_+a_2\right)&=&0,\\
\left(-\ii\omega+\tau_{\mathrm{mc}}^{-1}\right)a_2+
\frac{\ii v_F}{2}q_-a_1&=&0,
\end{eqnarray}
with conjugate equations for $m<0$.  The stress mode is massive on hydrodynamic time scales and can be eliminated algebraically,
\begin{equation}
a_2=-\frac{\ii v_F}{2}
\frac{q_-}{-\ii\omega+\tau_{\mathrm{mc}}^{-1}}a_1.
\end{equation}
Substituting back into the $m=1$ equation gives
\begin{equation}
\left[-\ii\omega+\frac{v_F^2}{4}
\frac{q^2}{-\ii\omega+\tau_{\mathrm{mc}}^{-1}}\right]a_1
+\frac{\ii v_F}{2}q_-a_0=0.
\end{equation}
The coefficient of $q^2a_1$ defines the frequency-dependent kinematic viscosity,
\begin{equation}
\nu_0(\omega)=\frac{v_F^2}{4}\frac{1}{\tau_{\mathrm{mc}}^{-1}-\ii\omega}\to\frac{v_F^2}{4}\tau_{\mathrm{mc}},
\end{equation}
in the dc limit.


\section{Schwinger--Keldysh formulation and random friction model}
\label{sec:section_3}

The clean modes above can be written either as equations of motion or as a real-time Schwinger--Keldysh (SK) field theory.  We use the latter because disorder averaging and interference ladders require causal response functions and equal-time fluctuations in the same language.  No new hydrodynamic assumption is being made; the SK formulation is the real-time bookkeeping that keeps retarded self-energies, Keldysh correlators, and disorder vertices tied to the same action. Below we introduce the SK machinery and its application to the clean hydrodynamic problem.

\subsection{Closed time contour and classical/quantum fields}

In the SK formalism, the system dynamically evolves along a contour $\mathcal{T}$ propagating from $t=-\infty$ to $t=\infty$ (forward) and back to $t=-\infty$ (backward). This closed time contour doubles each hydrodynamic field, each doubler propagating along each path,
\begin{equation}
a_{m,+}(t,\bm x),\quad a_{m,-}(t,\bm x),
\end{equation}
where the two copies refer to the forward and backward branches of the density-matrix evolution.  Physical configurations lie on the diagonal $a_{m,+}=a_{m,-}$.  We rotate to the classical/quantum basis \cite{Kamenev23,Kamenev99},
\begin{equation}
a_{m,\cl}=\frac{a_{m,+}+a_{m,-}}{2},
\quad
 a_{m,\q}=a_{m,+}-a_{m,-}.
\label{eq:cl_q_rotation_harmonics}
\end{equation}
Below we will move from the $a_{\pm1}$ modes to the velocity field $\bm u$, with the same convention,
\begin{equation}
\bm u_\cl=\frac{\bm u_++\bm u_-}{2},
\quad
\bm u_\q=\bm u_+-\bm u_- .
\label{eq:cl_q_rotation_velocity}
\end{equation}
The field $a_\cl$ is the physical hydrodynamic fluctuation, while $a_\q$ is the response field.  Setting all quantum fields to zero puts the two branches on top of each other; the generating functional must then be normalized to unity.  This causal normalization is the basic reason why every allowed SK action vanishes when the quantum fields are set to zero.

\subsection{Causal quadratic action}

For a linear hydrodynamic theory the clean equations of motion are encoded in a quadratic SK action.  Let $a$ denote collectively the retained harmonics.  If $L_0$ is the linear operator whose equation $L_0a_\cl=0$ reproduces the clean retarded hydrodynamic equations, the most general Gaussian action compatible with causality has the form
\begin{equation}
S_0=\int_{\omega,\bm q}
\left[
 a_\q^\dagger L_0 a_\cl
 +a_\cl^\dagger L_0^\dagger a_\q
 +\ii a_\q^\dagger N_0 a_\q
\right].
\label{eq:clean_SK_action}
\end{equation}
The first term imposes the retarded hydrodynamic equation on the classical field, the second is the advanced conjugate, and the last term contains the noise kernel $N_0$.  The absence of an $a_\cl^\dagger a_\cl$ term is the SK statement of causality and normalization: the action vanishes when all quantum fields are set to zero.

With the convention of Eq.~\eqref{eq:clean_SK_action}, the retarded and advanced propagators are
\begin{equation}
G^R=L_0^{-1},
\quad
G^A=(G^R)^\dagger .
\label{eq:retarded_advanced_from_L0}
\end{equation}
The Keldysh component is fixed in equilibrium by the fluctuation--dissipation relation
\begin{equation}
G^K(\omega)=\left(G^R-G^A\right)
\coth\left(\frac{\omega}{2T}\right),
\label{eq:KMS_relation_general}
\end{equation}
which becomes
\begin{equation}
G^K(\omega)\simeq \frac{2T}{\omega}\left(G^R-G^A\right)
\label{eq:classical_Keldysh_general}
\end{equation}
in the classical hydrodynamic regime $|\omega|\ll T$.  This is the only part of the noise kernel $N_0$ that is needed below.  In the one-loop retarded self-energies, the external legs are a quantum field and a classical field; the internal collective line is the Keldysh correlator.  This is why the SCBA and stress-sector diagrams reduce to equal-time fluctuation integrals.

\subsection{Random-friction coupling}

\begin{figure*}[t]
    \centering
    \includegraphics[width=0.82\textwidth]{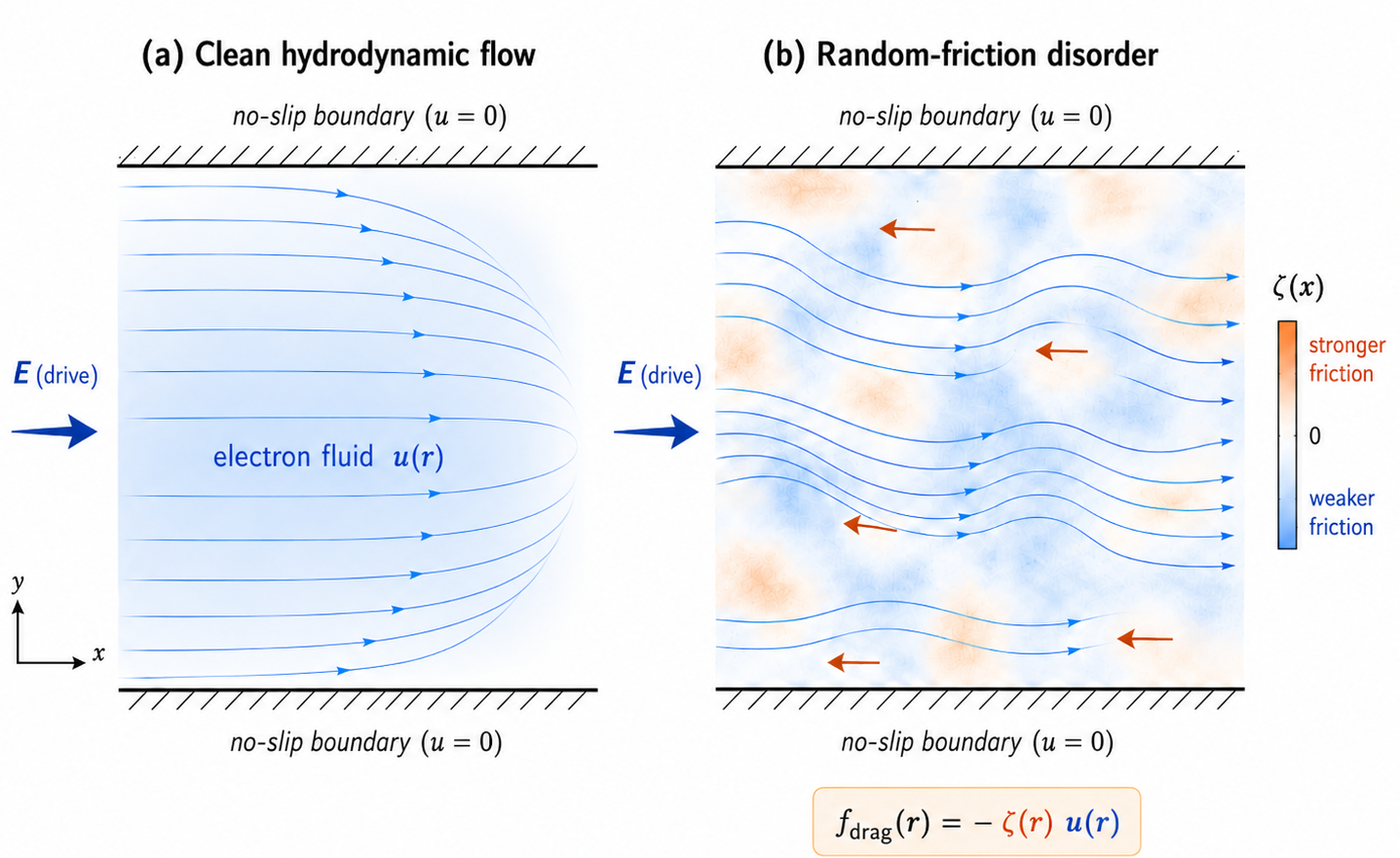}
    \caption{
    Schematic illustration of the random-friction model.
    (a) In the clean hydrodynamic regime, a uniform driving field $\bm E$ produces a smooth electron-fluid velocity profile $u(\mathbf r)$ in a channel with no-slip boundaries.
    (b) In the presence of static inhomogeneous random friction, described by a spatially varying field $\zeta(\mathbf r)$, the local drag force
    $f_{\mathrm{drag}}(\mathbf r)=-\zeta(\mathbf r)\,u(\mathbf r)$
    opposes the flow and distorts the velocity streamlines.
    The color map represents the spatial variation of the friction strength, with stronger-friction regions producing a larger local suppression of the flow.
    This disorder realization provides the minimal hydrodynamic mechanism used in the Schwinger--Keldysh treatment to generate momentum relaxation and construct the hydrodynamic Cooperon.
    }
    \label{fig:random_friction_scheme}
\end{figure*}

We now perturb the clean hydrodynamic fixed point by a static quenched random-friction field $\zeta(\bm x)$, interpreted as a local momentum-relaxing coefficient.  
The physical content of this minimal disorder model to be considered in this work is illustrated in Fig.~\ref{fig:random_friction_scheme}: static spatial variations of the local drag coefficient distort the hydrodynamic velocity field and provide the weak momentum-relaxing perturbation used below.

On the closed time contour the action on the backward branch enters with the opposite sign, so the disorder contribution is
\begin{equation}
S_{\mathrm{dis}}[\zeta]
= - \frac{1}{2}\int \dd t\,\dd^2x\,\zeta(\bm x)
\left[\bm u_+^2(t,\bm x)-\bm u_-^2(t,\bm x)\right].
\label{eq:Sdis_pm}
\end{equation}
Using Eq.~\eqref{eq:cl_q_rotation_velocity}, this becomes
\begin{equation}
S_{\mathrm{dis}}[\zeta]
= -\int \dd t\,\dd^2x\,\zeta(\bm x)
\bm u_\cl(t,\bm x)\cdot\bm u_\q(t,\bm x).
\label{eq:Sdis_clq}
\end{equation}
Thus random friction couples to one classical and one quantum velocity field.  This structure is required: the disorder vertex also vanishes in the physical limit $\bm u_\q=0$, so it does not spoil the SK normalization.

\subsection{Gaussian disorder average}

The disorder ensemble is Gaussian, static, and short-ranged ($\langle ...\rangle$ means averaging over disorder in the functional integral),
\begin{equation}
\langle \zeta(\bm x)\rangle=0,
\quad
\langle \zeta(\bm x)\zeta(\bm x')\rangle
=\Delta\,\delta^{(2)}(\bm x-\bm x').
\label{eq:disorder_statistics}
\end{equation}
The delta-correlated form is an effective long-wavelength description; loop momenta are restricted to the hydrodynamic window by the cutoff $\Lambda\sim\ell_h^{-1}=\ell_{\mathrm{mc}}^{-1}$.

Defining
\begin{equation}
J(\bm x)\equiv\int\dd t\,
\bm u_\cl(t,\bm x)\cdot\bm u_\q(t,\bm x),
\label{eq:J_disorder_source}
\end{equation}
the disorder average is a local Gaussian integral,
\begin{eqnarray}
\left\langle Z\right\rangle
=\int D\zeta\,\exp\Biggl[-\int\dd^2x\,\left(\frac{\zeta^2(\bm x)}{2\Delta}
-\ii\,\zeta(\bm x)J(\bm x)\right)\Biggr].
\label{eq:disorder_average_generating_functional}
\end{eqnarray}
Performing this integral gives
\begin{equation}
\left\langle Z\right\rangle
\propto
\exp\left[-\frac{\Delta}{2}\int\dd^2x\,J^2(\bm x)\right].
\label{eq:disorder_average_result}
\end{equation}
Equivalently, the disorder average produces the exact SK interaction
\begin{equation}
S_{\mathrm{int}}
=\frac{\ii\Delta}{2}\int\dd^2x
\left[\int\dd t\,
\bm u_\cl(t,\bm x)\cdot\bm u_\q(t,\bm x)
\right]^2.
\label{eq:Sint_disorder}
\end{equation}
The interaction is local in space but nonlocal in time, reflecting the quenched nature of the disorder.  Equation~\eqref{eq:Sint_disorder} is the vertex used below to generate the SCBA momentum-relaxing self-energy and the hydrodynamic Cooperon ladder.


\section{Self-consistent Born approximation (SCBA) for momentum relaxation}
\label{sec:section_4}

The clean fixed point has no momentum-relaxing term in the vector sector $m=\pm1$.  The static random-friction field generates such a rate only after disorder averaging.  Because the Cooperon ladder below is projected onto the transverse shear channel, we make the same projection in the SCBA self-energy.  The rate derived here is the disorder-induced damping of the transverse momentum mode that enters the Cooperon.  A full vector SCBA would add a longitudinal contribution and change nonuniversal prefactors, but it is not needed for the shear-channel calculation pursued in this paper.

\subsection{Transverse self-energy and SCBA equation}

The transverse dressed propagator is defined by
\begin{equation}
\left(G^R_\perp\right)^{-1}
=
\left(G^R_{\perp,0}\right)^{-1}+M^R_\perp,
\end{equation}
after using Eqs.~(\ref{eq:mode_equations}), with the sign convention that a positive real part of $M^R_\perp$ produces damping.  We define the momentum relaxing time as
\begin{equation}
\tau_{\mathrm{mr}}^{-1}
\equiv
\Re M^R_\perp(\omega\to0,\bm q=0).
\end{equation}
The quartic Schwinger--Keldysh vertex in Eq.~\eqref{eq:Sint_disorder} gives the standard one-loop self-energy
\begin{equation}
M^R_\perp(0)
= -\frac{\ii\Delta}{2}
\int\frac{\dd^2q}{(2\pi)^2}
\int\frac{\dd\omega}{2\pi}\,
G^K_\perp(\omega,\bm q) .
\label{eq:MR_Keldysh_loop}
\end{equation}
The SCBA consists of evaluating this internal line with the disorder-broadened transverse hydrodynamic propagator,
\begin{equation}
G^R_\perp(\omega,\bm q)
=
\frac{1}{\chi_P[-\ii\omega+\tau_{\mathrm{mr}}^{-1}+\nu q^2]}.
\label{eq:transverse_prop_SCBA}
\end{equation}
In equilibrium the Keldysh component is fixed by the KMS relation.  In the classical hydrodynamic regime $|\omega|\ll T$,
\begin{equation}
G^K_\perp(\omega,\bm q)
=
\frac{4\ii T}{\chi_P\{\omega^2+[\tau_{\mathrm{mr}}^{-1}+\nu q^2]^2\}},
\end{equation}
and therefore
\begin{equation}
\int\frac{\dd\omega}{2\pi}G^K_\perp(\omega,\bm q)
=
\frac{2\ii T}{\chi_P(\tau_{\mathrm{mr}}^{-1}+\nu q^2)}.
\end{equation}
Substitution into Eq.~\eqref{eq:MR_Keldysh_loop} gives the SCBA equation
\begin{equation}
\tau_{\mathrm{mr}}^{-1}
=
\frac{\Delta T}{\chi_P}
\int_{|\bm q|<\Lambda}
\frac{\dd^2q}{(2\pi)^2}
\frac{1}{\tau_{\mathrm{mr}}^{-1}+\nu q^2},
\label{eq:SCBA_transverse_integral}
\end{equation}
where $\Lambda\sim\ell_h^{-1}=\ell_{\mathrm{mc}}^{-1}$ is the hydrodynamic ultraviolet cutoff. The integral gives the self--consistent expression,
\begin{equation}
\tau_{\mathrm{mr}}^{-1}
=
\frac{\Delta T}{4\pi\chi_P\nu}
\ln\!\left(1+\frac{\nu\Lambda^2}{\tau_{\mathrm{mr}}^{-1}}\right).
\label{eq:SCBA_transverse_log}
\end{equation}
This equation makes explicit the role of the disorder-generated damping as the infrared regulator of the transverse hydrodynamic mode $m=\pm1$.  

\subsection{Weak-relaxation limit}

Equation \eqref{eq:SCBA_transverse_log} is an implicit equation for $\tau^{-1}_{\mathrm{mr}}$. In the weak-momentum-relaxation window relevant for the Cooperon pole,
\begin{equation}
\tau_{\mathrm{mr}}^{-1}\ll \nu\Lambda^2,
\label{eq:weak_mr_window_SCBA}
\end{equation}
it reduces to
\begin{equation}
\tau_{\mathrm{mr}}^{-1}
=
\frac{\Delta T}{4\pi\chi_P\nu}
\ln\!\left(\frac{\nu\Lambda^2}{\tau_{\mathrm{mr}}^{-1}}\right),
\label{eq:SCBA_single_log_effective}
\end{equation}
equation that can be solved explicitly,
\begin{equation}
\tau_{\mathrm{mr}}^{-1}(T)
=
\frac{\Delta T}{4\pi\chi_P\nu}
W_0\!\left(\frac{4\pi\chi_P \nu^2(T)\Lambda^2}{\Delta T}\right),
\label{eq:SCBA_Lambert_solution}
\end{equation}
where $W_0$ is the principal Lambert branch.  For $4\pi\chi_P\nu^2(T)/\Delta T\gg1$, the leading behavior is
\begin{equation}
\tau_{\mathrm{mr}}^{-1}(T)
\simeq
\frac{\Delta T}{4\pi\chi_P\nu(T)}
\ln\!\left[
\frac{4\pi\chi_P\nu^2(T)\Lambda^2}{\Delta T}
\right],
\label{eq:SCBA_leading_solution}
\end{equation}
up to subleading logarithmic corrections.  Thus the two-dimensional transverse hydrodynamic mode gives an infrared logarithm, cut off self-consistently by the disorder-generated momentum relaxation rate.

The longitudinal velocity sector can be included in a full vector SCBA, but the subsequent ladder and stress self-energy are projected onto the transverse shear channel.  Keeping the SCBA and Cooperon projections aligned avoids introducing a longitudinal sound-channel contribution whose main effect would be to alter nonuniversal prefactors rather than the Ward-identity structure of the viscosity correction.


\section{Hydrodynamic Cooperon ladder and infrared pole}
\label{sec:section_5}

The SCBA rate broadens the transverse momentum mode and supplies the infrared regulator for the maximally crossed ladder.  The resulting object is not a fermionic Cooperon; it is a collective interference mode built from hydrodynamic shear propagators.  We keep the transverse projection throughout.  Longitudinal fluctuations could be added to the momentum-relaxation rate, or approximated by the replacement $\nu\to\nu_L$, but in a compressible fluid they mix with density and lead to a matrix sound-channel ladder.  For the viscosity correction considered here, the scalar transverse Cooperon is the minimal shear-channel object.

The normalized transverse retarded and advanced propagators are
\begin{eqnarray}
\label{eq:GRGA_perp_hydro}
G^{R/A}_{\perp}(\omega,\bm q)&=&\frac{1}{\mp\ii\omega+\tau_{\mathrm{mr}}^{-1}+\nu q^2}.
\end{eqnarray}
The physical velocity correlator contains an additional factor $\chi_P^{-1}$, which we absorb into the definition of the disorder rung in the ladder channel.  This convention keeps the Cooperon algebra independent of the normalization chosen for the hydrodynamic velocity field.

\subsection{Retarded--advanced bubble}

The elementary kernel of the maximally crossed ladder is the transverse retarded--advanced bubble ($\omega_{\pm}=\omega\pm\Omega/2$, $\bm q_{\pm}=\bm q\pm \bm Q/2$)
\begin{eqnarray}
\label{eq:Pi_def}
\Pi_{\perp}(\Omega,\bm Q)
&\equiv&
\int\frac{\dd\omega}{2\pi}\int\frac{\dd^2 q}{(2\pi)^2}\,
G^R_{\perp}\!\left(\omega_+,\bm q_+\right)
\nonumber\\
&\times&
G^A_{\perp}\!\left(\omega_-,\bm q_-\right).
\end{eqnarray}
The nontrivial step is the frequency integral.  Defining
\begin{equation}
E_{\pm}=\tau_{\mathrm{mr}}^{-1}+\nu\left|\bm q\pm\frac{\bm Q}{2}\right|^2,
\end{equation}
the retarded and advanced poles lie in opposite half-planes, and the residue evaluation gives
\begin{eqnarray}
&\int\frac{\dd\omega}{2\pi}
\frac{1}{\left[E_+-\ii\left(\omega+\frac{\Omega}{2}\right)\right]
\left[E_-+\ii\left(\omega-\frac{\Omega}{2}\right)\right]}=\frac{1}{E_++E_- -\ii\Omega}.
\label{eq:freq_integral_result}
\end{eqnarray}
This is the hydrodynamic analogue of the standard retarded/advanced enhancement: the two poles pinch the real axis when the damping scale is small.
Using
\begin{equation}
\left|\bm q+\frac{\bm Q}{2}\right|^2+\left|\bm q-\frac{\bm Q}{2}\right|^2
=2q^2+\frac{Q^2}{2},
\end{equation}
one obtains
\begin{equation}
\Pi_{\perp}(\Omega,\bm Q)
=
\frac{1}{2}
\int\frac{\dd^2q}{(2\pi)^2}
\frac{1}{\tau_{\mathrm{mr}}^{-1}+\nu q^2+\frac{\nu Q^2}{4}-\ii\frac{\Omega}{2}}.
\label{eq:Pi_reduced}
\end{equation}
The factor $1/2$ and the combination $\nu Q^2/4-\ii\Omega/2$ are important: they fix the diffusion constant of the normalized Cooperon pole to be one half of the shear momentum diffusion constant.

The hydrodynamic theory is cut off at momenta of order
\begin{equation}
\Lambda\sim \ell_h^{-1}=\ell_{\mathrm{mc}}^{-1}.
\end{equation}
At finite external $Q$, imposing the cutoff on $|\bm q|$ rather than on the two shifted momenta $|\bm q\pm\bm Q/2|$ changes only nonuniversal boundary terms in the small-$Q$ expansion.  Introducing
\begin{equation}
\Gamma(\Omega,Q)
\equiv
\tau_{\mathrm{mr}}^{-1}+\frac{\nu Q^2}{4}-\ii\frac{\Omega}{2},
\label{eq:Gamma_def}
\end{equation}
the cutoff-regulated bubble is
\begin{eqnarray}
\Pi_{\perp}(\Omega,Q)
&=&
\frac{1}{2}\int_{|\bm q|<\Lambda}\frac{\dd^2q}{(2\pi)^2}
\frac{1}{\nu q^2+\Gamma(\Omega,Q)}
\nonumber\\
&=&
\frac{1}{8\pi\nu}
\ln\left[
\frac{\nu\Lambda^2+\Gamma(\Omega,Q)}{\Gamma(\Omega,Q)}
\right].
\label{eq:Pi_log_final}
\end{eqnarray}
This logarithm is the origin of the infrared sensitivity of the hydrodynamic Cooperon.  Its ultraviolet-dependent part is not universal; it parametrizes matching to local hydrodynamic coefficients at the cutoff scale.

\subsection{Ladder resummation}

\begin{figure}[t]
    \centering
    \includegraphics[width=\columnwidth]{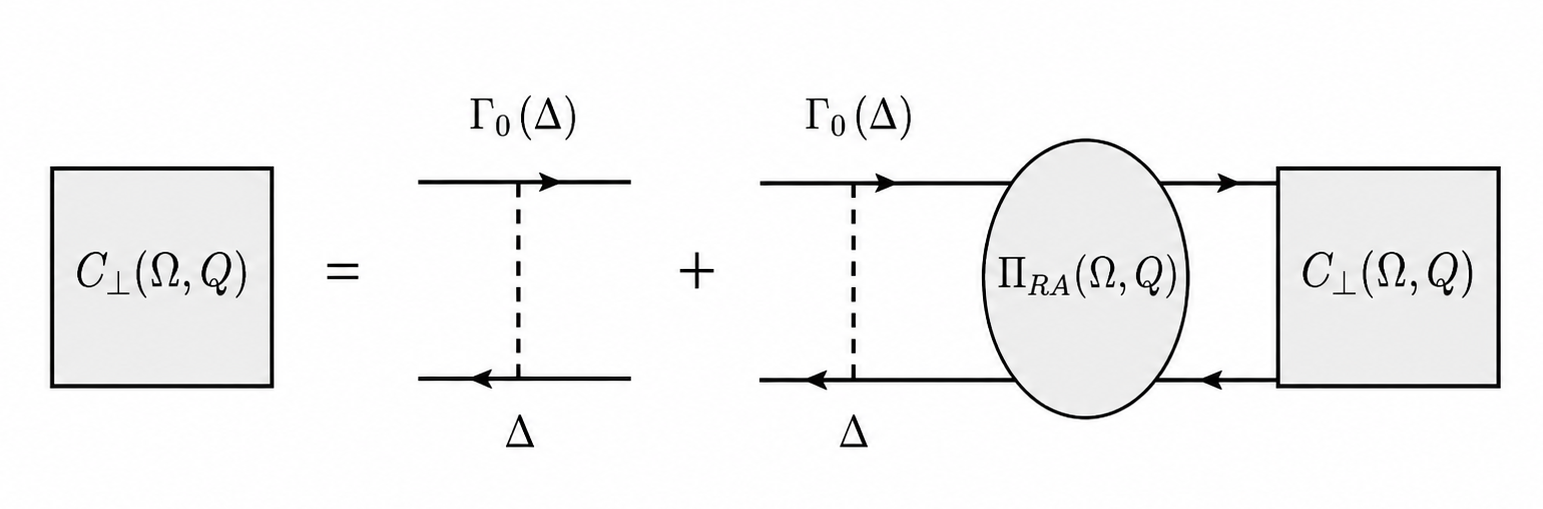}
    \caption{
    Diagrammatic Bethe--Salpeter equation for the transverse hydrodynamic Cooperon
    $C_{\perp}(\Omega,Q)$.
    The gray boxes denote the full Cooperon, the dashed vertical line is the
    random-friction disorder correlator $\Delta$, and $\Gamma_{0}(\Delta)$
    is the bare rung. The bubble $\Pi_{RA}(\Omega,Q)$ is built from one
    retarded and one advanced transverse hydrodynamic propagator.
    The second term contains one explicit disorder insertion; the Cooperon box
    on the right represents the remaining resummed ladder.
    }
    \label{fig:bethe_salpeter_cooperon}
\end{figure}

The corresponding Bethe--Salpeter equation is shown diagrammatically in Fig.~\ref{fig:bethe_salpeter_cooperon}. The algebraic structure is the same as in the diffusive electronic problem, but it is important to remember that the ladder is built from collective transverse hydrodynamic propagators in Eq.~(\ref{eq:GRGA_perp_hydro}), not from microscopic fermionic Green's functions.

The quartic interaction generated by disorder averaging supplies the bare rung of the maximally crossed ladder.  In the normalized transverse convention used here we denote this rung by
\begin{equation}
\Gamma_0\equiv \Delta,
\end{equation}
with the understanding that susceptibility factors have been absorbed into $\Gamma_0$.  The Bethe--Salpeter series is geometric,
\begin{equation}
C_{\perp}(\Omega,Q)
=
\Gamma_0
\left[1+\Gamma_0\Pi_{\perp}+(\Gamma_0\Pi_{\perp})^2+\cdots\right],
\end{equation}
and therefore
\begin{equation}
C_{\perp}(\Omega,Q)
=
\frac{\Gamma_0}{1-\Gamma_0\Pi_{\perp}(\Omega,Q)}.
\label{eq:Cooperon_solution}
\end{equation}
Equivalently,
\begin{equation}
C_{\perp}^{-1}(\Omega,Q)
=
\Gamma_0^{-1}-\Pi_{\perp}(\Omega,Q).
\label{eq:Cinv_def}
\end{equation}
Using Eq.~\eqref{eq:Pi_log_final}, the cutoff-regulated inverse Cooperon is
\begin{equation}
C_{\perp}^{-1}(\Omega,Q)
=
\Delta^{-1}
-
\frac{1}{8\pi\nu}
\ln\left[
\frac{\nu\Lambda^2+\Gamma(\Omega,Q)}{\Gamma(\Omega,Q)}
\right].
\label{eq:Cinv_explicit}
\end{equation}
In the next section we expand this expression at small $\Omega$ and $Q$ to extract the infrared pole parameters.  The condition that the zero-frequency, zero-momentum denominator remain positive is the Cooperon-channel stability condition.


\subsection{Infrared Cooperon pole}

The logarithmic ladder obtained in Eq.~(\ref{eq:Cinv_explicit}) contains both nonuniversal short-distance information and the infrared pole that will enter the stress self-energy.  We isolate the latter by expanding the inverse Cooperon for small center-of-mass frequency and momentum.  The starting point is
\begin{equation}
C_\perp^{-1}(\Omega,Q)
=
\Delta^{-1}
-\frac{1}{8\pi\nu}
\ln\!\left(\frac{\nu\Lambda^2+\Gamma(\Omega,Q)}{\Gamma(\Omega,Q)}\right),
\label{eq:Cinv_exact}
\end{equation}
with
\begin{equation}
\Gamma(\Omega,Q)
\equiv
\tau_{\mathrm{mr}}^{-1}+\frac{\nu Q^2}{4}-\ii\frac{\Omega}{2}.
\label{eq:Gamma_def_VI}
\end{equation}
The pole expansion is defined by
\begin{eqnarray}
C_\perp^{-1}(\Omega,Q)
&=&
C_0^{-1}(T)+D_C(T)Q^2-\ii Z_C(T)\Omega
\nonumber\\
&+&O(Q^4,\Omega^2,Q^2\Omega).
\label{eq:Cinv_lowE_def}
\end{eqnarray}
It is controlled for $|\Omega|\ll\tau_{\mathrm{mr}}^{-1}$ and $\nu Q^2\ll\tau_{\mathrm{mr}}^{-1}$, in addition to the hydrodynamic cutoff condition $Q\ll\Lambda$.  The elementary differentiation of Eq.~\eqref{eq:Cinv_exact} is given in Appendix~\ref{app:zero-field-pole}.  The resulting finite-cutoff coefficients are
\begin{eqnarray}
C_0^{-1}(T) &=&
\Delta^{-1}
-\frac{1}{8\pi\nu}
\ln\!\left(1+\frac{\nu\Lambda^2}{\tau_{\mathrm{mr}}^{-1}}\right),
\label{eq:C0inv}
\\
D_C(T)
&=&
\frac{1}{32\pi}
\left[
\frac{1}{\tau_{\mathrm{mr}}^{-1}}
-
\frac{1}{\tau_{\mathrm{mr}}^{-1}+\nu\Lambda^2}
\right],
\label{eq:DC_final}
\\
Z_C(T)
&=&
\frac{1}{16\pi\nu}
\left[
\frac{1}{\tau_{\mathrm{mr}}^{-1}}
-
\frac{1}{\tau_{\mathrm{mr}}^{-1}+\nu\Lambda^2}
\right].
\label{eq:ZC_final}
\end{eqnarray}
The ratio of the last two coefficients is independent of the ultraviolet regularization within the scalar transverse ladder,
\begin{equation}
\frac{D_C}{Z_C}=\frac{\nu}{2}.
\label{eq:DoverZ_zero}
\end{equation}
Thus the normalized Cooperon pole can be written as
\begin{equation}
C_\perp^R(\Omega,Q)
=
\frac{Z_C^{-1}}
{-\ii\Omega+\tau_C^{-1}(T)+D_C^{\rm hyd}(T)Q^2},
\label{eq:C_norm_zero}
\end{equation}
where
\begin{equation}
\tau_C^{-1}(T)=\frac{C_0^{-1}(T)}{Z_C(T)},
\quad
D_C^{\rm hyd}(T)=\frac{D_C(T)}{Z_C(T)}=\frac{\nu(T)}{2}.
\label{eq:C_pole_params_zero}
\end{equation}
The equality $D_C^{\rm hyd}=\nu/2$ is the hydrodynamic content of the pole: the Cooperon center-of-mass mode diffuses with one half of the shear momentum diffusivity.

In the weak-damping logarithmic regime,
\begin{equation}
\nu\Lambda^2\gg\tau_{\mathrm{mr}}^{-1},
\end{equation}
the finite-cutoff terms in Eqs.~\eqref{eq:DC_final} and \eqref{eq:ZC_final} are parametrically small, and
\begin{eqnarray}
D_C(T)&\simeq& \frac{1}{32\pi}\frac{1}{\tau_{\mathrm{mr}}^{-1}(T)},
\\
Z_C(T)&\simeq& \frac{1}{16\pi\nu(T)}\frac{1}{\tau_{\mathrm{mr}}^{-1}(T)}.
\label{eq:DCZC_weak}
\end{eqnarray}
The pole theory is meaningful only in the stable Cooperon regime.  Without additional dephasing this requires
\begin{equation}
C_0^{-1}(T)>0.
\label{eq:stability}
\end{equation}
If this positivity condition fails, the scalar Cooperon pole no longer represents a perturbative hydrodynamic correction; rather, it signals that the maximally crossed channel must be reorganized beyond the present EFT.

\section{Stress-sector self-energy and viscosity renormalization}
\label{sec:section_6}

The transverse Cooperon can affect observables only in channels allowed by the hydrodynamic Ward identities.  A uniform correction to the scalar density or vector momentum sector would give a dissipative mass to a conserved hydrodynamic variable, so it is forbidden at $(\omega,\bm q)=(0,\bm 0)$ in the clean limit. The first available channel is the nonconserved spin-two stress sector.  Its self-energy renormalizes the stress relaxation rate and, after the stress mode is eliminated, the shear viscosity.

Once disorder is present, momentum is no longer an exactly conserved quantity, and quantum corrections can in principle renormalize the disorder-induced momentum-relaxation rate $\tau_{\rm mr}^{-1}$.  Such terms dress the explicit translation-breaking operator and are therefore background renormalizations of the $m=\pm1$ sector.  The statement made here is different: at the clean hydrodynamic fixed point a uniform interference self-energy is forbidden in the conserved scalar and vector sectors, so the first intrinsic hydrodynamic transport coefficient that can
receive a Cooperon correction is the spin-two stress relaxation rate.

This projection is the Schwinger--Keldysh implementation of the Ward-identity argument formulated in the companion Letter.  There, the same constraint is expressed directly in the angular-harmonic language: at the clean hydrodynamic fixed point the $q=0$ equations for the conserved $m=0,\pm1$ sectors require
$\partial_t a_0=\partial_t a_{\pm1}=0$, and therefore force any uniform dissipative interference correction $\delta\Gamma_0,\delta\Gamma_{\pm1}$ to vanish.  The present work starts from that conservation-law projection and derives the corresponding Schwinger--Keldysh Cooperon self-energy explicitly.

The stress--momentum coupling follows directly from the $m=2$ equation in the Navier--Stokes ($m=\{0,\pm 1,\pm2\}$) hierarchy,
\begin{equation}
(\partial_t+\tau_{\mathrm{mc}}^{-1})a_2+\frac{v_F}{2}\partial_-a_1=0.
\label{eq:m2_eom_realspace}
\end{equation}
Using $u_-=v_F a_1$ and $u_+=v_F a_{-1}$, and writing $Q_\pm=Q_x\pm\ii Q_y$, this gives the stress--velocity vertices
\begin{equation}
V_{2\leftarrow u}(\bm Q)=\frac{\ii}{2}Q_-,
\quad
V_{u\leftarrow 2}(\bm Q)=\frac{\ii}{2}Q_+,
\label{eq:vertex_u}
\end{equation}
and hence
\begin{equation}
V_{2\leftarrow u}(\bm Q)V_{u\leftarrow 2}(\bm Q)=-\frac{Q^2}{4}.
\label{eq:vertex_product}
\end{equation}
This factor of $Q^2$ is the local signature of the stress projection.  It makes the uniform stress correction finite while preserving the protected conserved sectors.

The retarded stress self-energy $\Sigma_2^R(\Omega,\bm Q)$ is obtained by attaching external $(q,\mathrm{cl})$ stress legs and contracting the internal collective mode in the Keldysh component\cite{Kamenev99}.  At zero external frequency and momentum,
\begin{equation}
\Sigma_2^R(0,0)
=
\frac{\ii}{8}
\int\frac{\dd^2Q}{(2\pi)^2}
\int\frac{\dd\Omega}{2\pi}\,
Q^2 C_\perp^K(\Omega,Q),
\label{eq:SigmaR_simplified}
\end{equation}
where the sign and prefactor follow from Eq.~\eqref{eq:vertex_product}.  We define the positive correction to the stress relaxation eigenvalue by
\begin{equation}
\delta\Gamma_2(T)=-\Re\,\Sigma_2^R(0,0;T).
\label{eq:deltaGamma_def}
\end{equation}

The low-energy Cooperon entering Eq.~\eqref{eq:SigmaR_simplified} is
\begin{equation}
C_\perp^R(\Omega,Q)=
\frac{1}{C_0^{-1}(T)+D_C(T)Q^2-\ii Z_C(T)\Omega},
\label{eq:CR_low}
\end{equation}

Equilibrium fixes the Keldysh component through the KMS relation,
\begin{equation}
C_\perp^K(\Omega,Q)=
\bigl[C_\perp^R(\Omega,Q)-C_\perp^A(\Omega,Q)\bigr]
\coth\!\left(\frac{\Omega}{2T}\right).
\label{eq:CK_KMS}
\end{equation}
In the classical hydrodynamic regime, $|\Omega|\ll T$, the equal-time Cooperon fluctuation reduces to
\begin{equation}
\int\frac{\dd\Omega}{2\pi}\,C_\perp^K(\Omega,Q)
=
\frac{2\ii T}{C_0^{-1}(T)+D_C(T)Q^2}.
\label{eq:intCK_final}
\end{equation}
The cancellation of $Z_C$ in this expression is useful: the stress correction depends on the unnormalized mass and stiffness of the pole, but not on the dynamical normalization separately.  The short derivation of Eq.~\eqref{eq:intCK_final}, together with the radial momentum integral below, is recorded in Appendix~\ref{app:stress-self-energy}.

Combining Eqs.~\eqref{eq:SigmaR_simplified} and \eqref{eq:intCK_final} gives the central zero-field result in integral form,
\begin{equation}
\delta\Gamma_2(T)
=
\frac{T}{4}
\int_{|\bm Q|<\Lambda_C}
\frac{\dd^2Q}{(2\pi)^2}
\frac{Q^2}{C_0^{-1}(T)+D_C(T)Q^2}.
\label{eq:deltaGamma_integral}
\end{equation}
Here $\Lambda_C$ is the cutoff of the infrared Cooperon-pole approximation, not the microscopic hydrodynamic cutoff by itself.  The pole expansion requires
\begin{equation}
|\Omega|\ll\tau_{\mathrm{mr}}^{-1},
\quad
\nu Q^2\ll\tau_{\mathrm{mr}}^{-1},
\end{equation}
in addition to the hydrodynamic conditions $Q\ll\Lambda_h\sim\ell_{\mathrm{mc}}^{-1}$ and $|\Omega|\tau_{\mathrm{mc}}\ll1$.  Thus a controlled pole-level evaluation uses
\begin{equation}
\Lambda_C\lesssim
\min\left(\Lambda_h,\sqrt{\frac{\tau_{\mathrm{mr}}^{-1}}{\nu}}\right).
\label{eq:LambdaC_zero}
\end{equation}
Taking $\Lambda_C\sim\Lambda_h$ is a phenomenological extrapolation of the pole form across the full hydrodynamic window.  In that case the leading power-law piece below should be regarded as a local renormalization of the bare spin-two relaxation rate, while the logarithmic dependence on the Cooperon mass is the infrared-sensitive part.

The two-dimensional integral in Eq.~\eqref{eq:deltaGamma_integral} gives
\begin{eqnarray}
&&\delta\Gamma_2(T)
=
\frac{T\Lambda_C^2}{16\pi D_C(T)}
\nonumber\\
&-&
\frac{T C_0^{-1}(T)}{16\pi D_C^2(T)}
\ln\!\left[1+\frac{D_C(T)\Lambda_C^2}{C_0^{-1}(T)}\right].
\label{eq:deltaGamma_closed}
\end{eqnarray}
For $C_0^{-1}(T)>0$ and $D_C>0$ this correction is positive.  The renormalized stress relaxation rate is
\begin{equation}
\Gamma_\pi^{\mathrm{eff}}(T)=\tau_{\mathrm{mc}}^{-1}(T)+\delta\Gamma_2(T),
\label{eq:Gamma_pi_eff_zero}
\end{equation}
and the dc shear viscosity in the kinetic realization becomes
\begin{equation}
\nu_{\mathrm{eff}}(T)=
\frac{v_F^2}{4\Gamma_\pi^{\mathrm{eff}}(T)}.
\label{eq:nu_eff_zero}
\end{equation}
The hydrodynamic Cooperon thus increases the decay rate of the spin-two mode and lowers the viscosity.  In a narrow Gurzhi channel, where the viscous resistivity is proportional to the viscosity, this gives a negative correction to the viscous contribution to the resistance.

This completes the zero-field derivation.  The magnetic-field problem is not obtained by merely adding a dephasing rate to Eq.~\eqref{eq:DoverZ_zero}.  A perpendicular field first changes the clean hydrodynamic fixed point itself: the stress harmonics precess, the viscosity becomes a tensor, and the transverse Cooperon coefficients acquire field dependence.  Only after this clean magnetohydrodynamic structure is established does the orbital field act as a Cooperon cutoff through the gauge-covariant center-of-mass motion.  The remaining sections implement these two effects in that order.

\section{Magnetohydrodynamic fixed point and magnetic viscosities}
\label{sec:section_7}

We next consider the clean two--dimensional fluid in presence of a uniform perpendicular magnetic field.  The field does not first appear as a dephasing rate.  It changes the clean fixed point itself: the Lorentz force is reversible and belongs in the quadratic hydrodynamic kernel, whereas $\tau_{\mathrm{mr}}^{-1}$ is still generated only by disorder.  This ordering matters.  The magnetic field changes the bare propagators and the viscosity tensor before it cuts off the Cooperon orbitally.

In the kinetic representation the field enters through the angular precession term in the linearized Boltzmann equation,
\begin{equation}
\partial_t f+v_F\hat{\bm k}\cdot\nabla f
-\omega_c\partial_\theta f
=I_{\mathrm{mc}}[f]+ne\bm E/\chi_P,
\label{eq:mag_kinetic}
\end{equation}
with $\omega_c=eB/m^\ast$ and with $ne\bm E/\chi_P$ is an electric-field driving force.  Projection onto angular harmonics gives
\begin{eqnarray}
\partial_t a_m
&+&\frac{v_F}{2}\left(\partial_-a_{m-1}+\partial_+a_{m+1}\right)
-\ii m\omega_c a_m
\nonumber\\
&=&-\gamma_m^{(0)}a_m+s_m ,
\label{eq:mag_harmonic_general}
\end{eqnarray}
with $s_{\pm1}=\frac{n e}{\chi_P v_F}(E_x\mp\ii E_y)$ and $s_m=0$ for $m\neq\pm1$.  Thus the $m=\pm1$ momentum harmonics precess at the cyclotron frequency, while the $m=\pm2$ stress harmonics precess at twice that frequency.  Conservation laws are otherwise unchanged,
\begin{equation}
\gamma_0^{(0)}=\gamma_{\pm1}^{(0)}=0,
\quad
\gamma_{\pm2}^{(0)}=\tau_{\mathrm{mc}}^{-1},
\label{eq:mag_conservation_rates}
\end{equation}
so there is still no clean momentum-relaxing term in the vector sector.

The key difference from the zero-field calculation appears when the fast spin-two harmonics are projected out.  In zero field this step produced the scalar kinematic viscosity $\nu=v_F^2\tau_{\mathrm{mc}}/4$.  In a magnetic field the stress denominator is shifted by the spin-two precession frequency, and the viscosity becomes complex even at the dc limit,
\begin{eqnarray}
\nu_c^{(0)}(T,B)
&=&\frac{v_F^2}{4}\frac{1}{\tau_{\mathrm{mc}}^{-1}(T)-2\ii\omega_c}
\nonumber\\
&=&\nu_{xx}^{(0)}(T,B)+\ii\nu_{xy}^{(0)}(T,B).
\label{eq:complex_bare_mag_visc}
\end{eqnarray}
Equivalently,
\begin{eqnarray}
\nu_{xx}^{(0)}(T,B)&=&\frac{v_F^2}{4}
\frac{\tau_{\mathrm{mc}}^{-1}(T)}{\tau_{\mathrm{mc}}^{-2}(T)+4\omega_c^2},
\label{eq:bare_nu_xx_mag}\\
\nu_{xy}^{(0)}(T,B)&=&\frac{v_F^2}{4}
\frac{2\omega_c}{\tau_{\mathrm{mc}}^{-2}(T)+4\omega_c^2}.
\label{eq:bare_nu_xy_mag}
\end{eqnarray}
The dissipative longitudinal viscosity is suppressed by the classical Lorentz rotation of the stress tensor, while a reactive Hall-viscous component appears at the classical level~\cite{Avron98,Scaffidi17,Pellegrino17,Delacretaz17,Alekseev16,Alekseev20}.

In terms of macroscopic variables, the same reduction gives the clean magnetohydrodynamic equation
\begin{eqnarray}
\partial_t\bm u&+&\nabla\mu
=\frac{e}{m^\ast}\bm E+\omega_c(\bm u\times\hat{\bm z})
\nonumber\\
&+&\nu_{xx}^{(0)}(T,B)\nabla^2\bm u
-\nu_{xy}^{(0)}(T,B)\nabla^2(\bm u\times\hat{\bm z}).
\label{eq:clean_mag_hydro_eq}
\end{eqnarray}
The last term is the standard parity-odd viscous force in two-dimensional magnetohydrodynamics~\cite{Avron98,Pellegrino17,Delacretaz17,Berdyugin19}.  Here $\mu$ denotes the hydrodynamic scalar potential, normalized in the kinetic notation by $\mu=(v_F^2/2)a_0$, so that the pressure or chemical-potential force is written without reintroducing the angular harmonic $a_0$ into the macroscopic equation.
The auxiliary harmonic algebra leading from Eq.~\eqref{eq:mag_harmonic_general} to Eqs.~\eqref{eq:complex_bare_mag_visc} and \eqref{eq:clean_mag_hydro_eq} is recorded in Appendix~\ref{app:mag-harmonic-reduction}.  For the main argument the essential points are the following.  First, the magnetic field modifies the bare viscosity tensor through the spin-two sector.  Second, the Hall-viscous coefficient will enter the magnetic Cooperon through reactive shifts of the chiral hydrodynamic modes, while the longitudinal coefficient controls their damping.


\section{Magnetic SCBA and hydrodynamic Cooperon}
\label{sec:section_8}

The magnetic field affects the disorder problem in two logically distinct ways.  First, it changes the clean hydrodynamic propagators through the longitudinal and Hall viscosities derived in Sec.~\ref{sec:section_7}.  Second, it cuts off the interference ladder through the orbital motion of the Cooperon center of mass.  The first effect is a property of the magnetohydrodynamic fixed point, while the second is a genuine Cooperon dephasing mechanism.  This section treats the SCBA rate and the hydrodynamic Cooperon constructed from the magnetic clean fixed point~\cite{Alekseev16,Alekseev20,Avron98,Pellegrino17,Delacretaz17,Berdyugin19}; the orbital Cooperon mass is introduced in Sec.~\ref{sec:section_11}.

The random-friction vertex itself is unchanged by the field.  The disorder still couples locally to the velocity sector through the SK vertex defined in Sec.~\ref{sec:section_3}, and the magnetic field enters the SCBA and ladder problems only through the hydrodynamic propagators carried by the internal lines.  This separation follows the same quenched-disorder and SK logic used at zero field~\cite{Kamenev23,CrossleyGloriosoLiu17,GloriosoLiu18,Lee85,Vollhardt80b,Hershfield86}.

It is convenient to use the chiral velocity variables $u_-=v_Fa_1$ and $u_+=v_Fa_{-1}$.  After the stress modes have been integrated out, the SCBA-dressed retarded propagators may be written in the compact form
\begin{equation}
G_\sigma^R(\omega,q)=
\frac{1}{\chi_P}
\frac{1}{E_q(T,B)-\ii[\omega-\sigma\omega_c+
\sigma\nu_{xy}^{(0)}(T,B)q^2]},
\label{eq:mag_GR_chiral}
\end{equation}
where $\sigma=\pm1$ and $E_q(T,B)=\tau_{\mathrm{mr}}^{-1}(T,B)+\nu_{xx}^{(0)}(T,B)q^2$.  The real part of the denominator is controlled by the dissipative longitudinal magnetoviscosity, whereas the cyclotron and Hall-viscous terms are reactive frequency shifts.

The magnetic SCBA equation follows from the same equal-time Keldysh loop as at zero field.  Since the internal frequency is integrated over the hydrodynamic range, the uniform reactive shift can be removed by a change of variables; it does not by itself generate damping.  In the transverse/chiral approximation used throughout the Cooperon construction one obtains
\begin{eqnarray}
\label{eq:mag_SCBA}
&&\tau_{\mathrm{mr}}^{-1}(T,B)
=\frac{\Delta T}{4\pi\chi_P\nu_{xx}^{(0)}(T,B)}
\\\nonumber
&&\ln\left[1+
\frac{\nu_{xx}^{(0)}(T,B)\Lambda^2}{\tau_{\mathrm{mr}}^{-1}(T,B)}
\right].
\end{eqnarray}
This is the magnetic analogue of the zero-field transverse SCBA equation Eq.~(\ref{eq:SCBA_single_log_effective}).  Its field dependence is controlled by $\nu_{xx}^{(0)}(T,B)$, not by the Hall shift itself.  In the logarithmic regime $\nu_{xx}^{(0)}\Lambda^2\gg\tau_{\mathrm{mr}}^{-1}$, Eq.~\eqref{eq:mag_SCBA} gives the leading-log estimate $\tau_{\mathrm{mr}}^{-1}\simeq\gamma_B\ln(E_B/\gamma_B)$, with $\gamma_B=\Delta T/[4\pi\chi_P\nu_{xx}^{(0)}]$ and $E_B=\nu_{xx}^{(0)}\Lambda^2$.  Thus $\tau_{\mathrm{mr}}^{-1}$ scales approximately as $[\nu_{xx}^{(0)}]^{-1}$ up to the slowly varying logarithm ~\cite{Lee85,Vollhardt80b,Altshuler82,Hershfield86,Bergmann84}.

We now turn to the magnetic hydrodynamic Cooperon.  The ladder is built from a retarded and an advanced chiral propagator, exactly as in the zero-field retarded--advanced construction of Sec.~\ref{sec:section_5}.  The uniform cyclotron frequency cancels between the two lines.  The Hall-viscous shift, however, depends on momentum and therefore leaves a term linear in $\bm q\cdot\bm Q$.  After the internal frequency integral, the magnetic bubble reduces to
\begin{eqnarray}
\Pi_B(\Omega,\bm Q)
&=&\frac12\int\frac{\dd^2q}{(2\pi)^2}
\biggl[
E_q(T,B)+\frac{\nu_{xx}^{(0)}(T,B)}{4}Q^2
\nonumber\\
&-&\frac{\ii\Omega}{2}-\ii\nu_{xy}^{(0)}(T,B)\bm q\cdot\bm Q
\biggr]^{-1}.
\label{eq:mag_bubble_reduced}
\end{eqnarray}
This equation contains the central physical distinction between damping and interference.  The SCBA rate depends only on the dissipative denominator after the reactive shifts have been translated away.  The Cooperon pole also loses the uniform cyclotron frequency, but the Hall-viscous momentum dependence changes the spatial stiffness of the interference mode~\cite{Wu22,Avron98,Pellegrino17,Delacretaz17}.

Expanding Eq.~\eqref{eq:mag_bubble_reduced} at small frequency and momentum gives the magnetic Cooperon pole
\begin{equation}
C_B^{-1}(\Omega,Q)=
C_0^{-1}(T,B)+D_C(T,B)Q^2-
\ii Z_C(T,B)\Omega.
\label{eq:mag_C_inverse_pole}
\end{equation}
The explicit integral and finite-cutoff expressions for $C_0^{-1}$, $Z_C$, and $D_C$ are recorded in Appendix~\ref{app:mag-cooperon-coeffs}.  The only structural point needed in the main text is that $D_C$ contains the zero-field shear contribution continued to finite $B$ plus a new even-in-$B$ term proportional to $[\nu_{xy}^{(0)}]^2$.  The normalized diffusion constant of the magnetic Cooperon pole therefore takes the weak-damping form
\begin{eqnarray}
\label{eq:mag_Dhyd_weak}
D_C^{\mathrm{hyd}}(T,B)
&\equiv& \frac{D_C(T,B)}{Z_C(T,B)}
\\\nonumber
&\simeq&\frac{\nu_{xx}^{(0)}(T,B)}{2}
\left[1+\frac{[\nu_{xy}^{(0)}(T,B)]^2}{[\nu_{xx}^{(0)}(T,B)]^2}\right].
\end{eqnarray}
At $B=0$ this reduces to $D_C^{\mathrm{hyd}}=\nu/2$, as in Sec.~\ref{sec:section_5}.  For the simple kinetic magnetoviscosity of Sec.~\ref{sec:section_7}, the bracket compensates the Lorentz suppression of $\nu_{xx}^{(0)}$, giving $D_C^{\mathrm{hyd}}\simeq\nu_0(T)/2$ within the weak-damping approximation.  This does not mean that the magnetic field has produced a Cooperon mass.  It means that the classical Lorentz force has reshuffled the pole coefficients.  The genuine orbital mass of the Cooperon is a separate gauge-covariant effect, introduced next.


\section{Orbital Cooperon mass and magnetic stress correction}
\label{sec:section_9}

The magnetic coefficients obtained above are not, by themselves, the usual orbital dephasing of a maximally crossed interference mode.  Orbital dephasing is a separate gauge-covariant effect.  A Cooperon compares a trajectory with its time-reversed partner, so the phase accumulated around a closed loop is twice the single-particle Aharonov--Bohm phase.  Equivalently, the center-of-mass coordinate of the two-particle ladder behaves as an object of charge $2e$.  The long-wavelength implementation is the minimal, or Peierls, substitution in the center-of-mass momentum of the Cooperon pole.

This distinction matters because the Lorentz terms in Sec.~\ref{sec:section_8}   do not produce an infrared Cooperon mass by themselves: the uniform cyclotron shift cancels between the retarded and advanced propagators.  The mass comes from the noncommutativity of the covariant center-of-mass momenta in a magnetic field, which turns the Cooperon continuum into Landau levels.

The normalized magnetic Cooperon obtained in Sec.~\ref{sec:section_8} can be written as
\begin{eqnarray}
C_B^R(\Omega,Q)
&=&\frac{1}{Z_C(T,B)}
\left[
-\ii\Omega+\tau_C^{-1}(T,B)
\right.
\nonumber\\
&&\hspace{1.5cm}\left.
+D_C^{\mathrm{hyd}}(T,B)Q^2
\right]^{-1},
\nonumber\\
\tau_C^{-1}(T,B)&=&\frac{C_0^{-1}(T,B)}{Z_C(T,B)}.
\label{eq:CB_norm_sectionX}
\end{eqnarray}
Gauge covariance promotes the center-of-mass momentum to
\begin{equation}
\bm Q\rightarrow\bm\Pi\equiv -\ii\nabla-\frac{2e}{\hbar}\bm A .
\label{eq:peierls_substitution}
\end{equation}
For a uniform field $\bm B=B\hat{\bm z}$, the covariant Laplacian has the Landau-level spectrum
\begin{eqnarray}
\bm\Pi^2\psi_n&=&\lambda_n(B)\psi_n,
\nonumber\\
\lambda_n(B)&=&\frac{4e|B|}{\hbar}\left(n+\frac12\right),
\quad n=0,1,2,\ldots .
\label{eq:cooperon_LL_spectrum}
\end{eqnarray}
The orbital contribution to the normalized Cooperon denominator is
\begin{equation}
\tau_{B,n}^{-1}(T,B)=D_C^{\mathrm{hyd}}(T,B)\lambda_n(B).
\label{eq:tauBn_def}
\end{equation}
The long-time response is controlled by the lowest Landau level.  We define the corresponding
orbital magnetic dephasing rate by
\begin{eqnarray}
\label{eq:tauB_LLL}
&&\tau_B^{-1}(T,B)
\equiv\tau_{B,0}^{-1}(T,B)
\nonumber\\
&=&\frac{2e|B|}{\hbar}D_C^{\mathrm{hyd}}(T,B)
=\frac{2e|B|}{\hbar}\frac{D_C(T,B)}{Z_C(T,B)}.
\end{eqnarray}
It is important to mention here that the lowest Landau level controls the infrared structure of the full Landau-level tower, it is not a phenomenological dephasing ansatz.

Including an intrinsic dephasing, the unnormalized magnetic Cooperon mass becomes
\begin{eqnarray}
A_{\phi,B}(T,B)
&=&C_0^{-1}(T,B)+Z_C(T,B)\tau_\phi^{-1}(T)
\nonumber\\
&+&Z_C(T,B)\tau_B^{-1}(T,B).
\label{eq:AphiB_def}
\end{eqnarray}
Using Eq.~\eqref{eq:tauB_LLL}, this can be written in the particularly transparent form
\begin{eqnarray}
A_{\phi,B}(T,B)
&=&C_0^{-1}(T,B)+Z_C(T,B)\tau_\phi^{-1}(T)
\nonumber\\
&+&\frac{2e|B|}{\hbar}D_C(T,B).
\label{eq:AphiB_simplified}
\end{eqnarray}
The last term is the orbital mass in the unnormalized pole convention used for the stress
self-energy.

The projection of this magnetic Cooperon onto the stress sector is the same as at zero field.  The
magnetic field changes the stress propagator through the replacement
$\tau_{\mathrm{mc}}^{-1}\to\tau_{\mathrm{mc}}^{-1}-2\ii\omega_c$, but the stress--velocity vertex still contains a
single spatial gradient.  Consequently
\begin{equation}
V_{2\leftarrow u}(\bm Q)V_{u\leftarrow2}(\bm Q)=-\frac{Q^2}{4},
\label{eq:mag_stress_vertex_product}
\end{equation}
exactly as in the zero-field shear-channel calculation (\ref{eq:vertex_product}).  The infrared Cooperon entering the stress
self-energy is now
\begin{equation}
C_{\phi,B}^{R}(\Omega,Q)=
\frac{1}{A_{\phi,B}(T,B)+D_C(T,B)Q^2-\ii Z_C(T,B)\Omega}.
\label{eq:CphiB_pole}
\end{equation}
The KMS relation and the classical low-frequency limit again give the equal-time fluctuation
\begin{eqnarray}
\int\frac{\dd\Omega}{2\pi}\,C_{\phi,B}^{K}(\Omega,Q)
&=&\frac{2\ii T}{A_{\phi,B}(T,B)+D_C(T,B)Q^2}.
\label{eq:mag_CK_equal_time}
\end{eqnarray}
Thus the magnetic Cooperon correction to the spin-two relaxation rate is
\begin{eqnarray}
\label{eq:mag_deltaGamma_integral}
\delta\Gamma_2(T,B)
&=&\frac{T}{4}\int_{Q<\Lambda_C(B)}\frac{\dd^2Q}{(2\pi)^2}
\nonumber\\
&\times&
\frac{Q^2}{A_{\phi,B}(T,B)+D_C(T,B)Q^2}.
\end{eqnarray}
The pole cutoff must lie within both the hydrodynamic window and the magnetic Cooperon pole
regime.  Parametrically,
\begin{equation}
\Lambda_C(B)\lesssim
\min\left(\Lambda,\sqrt{\frac{\tau_{\mathrm{mr}}^{-1}(T,B)}{\nu_{xx}^{(0)}(T,B)}}\right),
\label{eq:mag_pole_cutoff}
\end{equation}
up to nonuniversal details of the matching between the pole theory and the full hydrodynamic
ladder.

The momentum integral has the same form as in the zero-field calculation, with the replacements
$C_0^{-1}\to A_{\phi,B}$ and $D_C\to D_C(T,B)$.  This gives
\begin{eqnarray}
\label{eq:mag_deltaGamma_closed}
&&\delta\Gamma_2(T,B)
=\frac{T\Lambda_C^2(B)}{16\pi D_C(T,B)}
\nonumber\\
&-&\frac{T A_{\phi,B}(T,B)}{16\pi D_C^2(T,B)}
\ln\left[1+\frac{D_C(T,B)\Lambda_C^2(B)}{A_{\phi,B}(T,B)}\right].
\end{eqnarray}
Equation~\eqref{eq:mag_deltaGamma_closed} makes explicit that the magnetic field suppresses the
interference correction in two ways.  First, the clean magnetic fixed point changes the coefficients
$D_C$, $Z_C$, $C_0^{-1}$, and $\tau_{\mathrm{mr}}^{-1}$ through $\nu_{xx}^{(0)}$ and $\nu_{xy}^{(0)}$.  Second,
the Peierls substitution adds the positive orbital mass $(2e|B|/\hbar)D_C$ to the unnormalized
Cooperon denominator.  It is the second effect that represents genuine orbital dephasing of the
maximally crossed ladder.

The zero-field limit is continuous.  As $B\to0$, $\nu_{xx}^{(0)}(T,B)$ approaches $\nu(T)$ in Eq.~(\ref{eq:nu_eff_zero}), $\nu_{xy}^{(0)}(T,B)\to0$, and $\tau_B^{-1}(T,B)\to0$, so that $D_C(T,B)$ and $Z_C(T,B)$ approach the zero-field results in Eq.~(\ref{eq:DCZC_weak}) and $A_{\phi,B}(T,B)$ approaches Eq.~(\ref{eq:C_norm_zero}).

Assuming the same infrared cutoff convention, Eq.~\eqref{eq:mag_deltaGamma_closed} reduces to
\begin{eqnarray}
&&\delta\Gamma_2(T,B\to0)
=\frac{T\Lambda_C^2}{16\pi D_C(T)} \\
&-&\frac{T A_\phi(T)}{16\pi D_C^2(T)}
\ln\left[1+\frac{D_C(T)\Lambda_C^2}{A_\phi(T)}\right],
\label{eq:mag_deltaGamma_zero_limit}
\end{eqnarray}
which is precisely the zero-field result in Eq.~(\ref{eq:deltaGamma_closed}).  In the purely orbital limit,
$C_0^{-1}\to0$ and $\tau_\phi^{-1}\to0$, Eq.~\eqref{eq:AphiB_simplified} gives
$A_B=(2e|B|/\hbar)D_C$, and Eq.~\eqref{eq:mag_deltaGamma_closed} reduces to
\begin{eqnarray}
&&\delta\Gamma_2^{\mathrm{orb}}(T,B)
=\frac{T}{16\pi D_C(T,B)}
\left[\Lambda_C^2(B)\right.\\
&&\quad\left.
-\frac{2e|B|}{\hbar}
\ln\left(1+\frac{\hbar\Lambda_C^2(B)}{2e|B|}\right)
\right].
\label{eq:mag_deltaGamma_orb}
\end{eqnarray}
This expression displays the physical effect most directly: a weak orbital field cuts off the
hydrodynamic Cooperon and continuously quenches its contribution to the spin-two relaxation rate.


\section{Experimental signatures}
\label{sec:section_10}

The preceding sections determine how the hydrodynamic Cooperon modifies the stress relaxation rate in a perpendicular magnetic field.  We now translate this result into experimentally visible magnetoviscous responses.  The purpose of this section is not to fit a specific material, but to separate the broad semiclassical magnetohydrodynamic background from the narrower coherence-sensitive structure produced by the stress-sector Cooperon.  The discussion below uses standard viscous-channel and Hall-viscosity phenomenology~\cite{Gurzhi63,deJong95,Jaggi91,Alekseev16,Alekseev20,Scaffidi17,Pellegrino17,Delacretaz17,Berdyugin19}.

At the clean magnetohydrodynamic level, the perpendicular field turns the scalar viscosity into a tensor.  In the kinetic Fermi-liquid notation used above, with $\Gamma_\pi(T)=\tau_{\rm mc}^{-1}(T)$, the two bare components are
\begin{equation}
\nu_{xx}^{(0)}(T,B)
=
\frac{v_F^2}{4}
\frac{\Gamma_\pi(T)}{\Gamma_\pi^2(T)+4\omega_c^2}.
\label{eq:bare_nuxx_exp}
\end{equation}
The corresponding Hall-viscous component is
\begin{equation}
\nu_{xy}^{(0)}(T,B)
=
\frac{v_F^2}{4}
\frac{2\omega_c}{\Gamma_\pi^2(T)+4\omega_c^2}.
\label{eq:bare_nuxy_exp}
\end{equation}
The factor $2\omega_c$ is the spin-two precession frequency of the stress sector, and Eqs.~\eqref{eq:bare_nuxx_exp} and \eqref{eq:bare_nuxy_exp} are the semiclassical reference against which the quantum correction should be compared~\cite{Alekseev16,Alekseev20,Avron98,Scaffidi17,Pellegrino17,Delacretaz17}.

The hydrodynamic Cooperon does not add a new conserved-mode self-energy.  Instead, it replaces the stress relaxation rate by $\Gamma_\pi^{\rm eff}(T,B)=\Gamma_\pi(T)+\delta\Gamma_2(T,B)$ in the same magnetoviscous tensor.  Since $\delta\Gamma_2(T,B)$ is largest near zero field and is suppressed by orbital Cooperon dephasing, this replacement produces a narrow low-field correction on top of the broader Lorentz suppression already present in the bare viscosity.  This is the central experimental distinction: the semiclassical background is controlled by stress precession~\cite{Lee85,Bergmann84,Altshuler82,Hershfield86,Alekseev16}, whereas the quantum feature is controlled by the magnetic cutoff of the Cooperon.

\subsection{Channel magnetoresistance}

For a strip of width $W$, the longitudinal channel resistivity is well represented by the viscous-channel expression
\begin{equation}
\rho_{\rm ch}(B,T)
=
\frac{\rho_{\rm mr}(B,T)}
{1-\frac{2D_v(B,T)}{W}\tanh\left[\frac{W}{2D_v(B,T)}\right]},
\label{eq:channel_resistance_full}
\end{equation}
where $D_v(B,T)=\sqrt{\nu_{xx}(B,T)\tau_{\rm mr}(B,T)}$.  The semiclassical curve follows from using $\nu_{xx}^{(0)}$, while the quantum-corrected curve follows from using $\nu_{xx}^{\rm eff}$.  In the deep Gurzhi limit, $W\ll D_v$, this expression simply reduces to
\begin{equation}
\rho_{\rm visc}(B,T)
\simeq
\frac{12m^\ast}{ne^2W^2}\,\nu_{xx}(B,T).
\label{eq:deep_gurzhi_resistance}
\end{equation}
Thus a reduction of the effective viscosity lowers the viscous resistance and raises the channel conductance~\cite{Gurzhi63,deJong95,Molenkamp94,Gurzhi95,Jaggi91,Alekseev16,Berdyugin19}.

This sign is opposite to the familiar weak-localization correction to the conductivity in a diffusive metal.  The difference is not the sign of the Cooperon itself, but the hydrodynamic projection enforced by conservation laws: charge and momentum remain protected, while the first allowed uniform correction is a positive correction to the nonconserved spin-two relaxation rate.  The resulting decrease of viscosity is therefore a hydrodynamic interference signature rather than a direct localization correction to charge diffusion~\cite{Lee85,Bergmann84,Altshuler82,Hershfield86,Wu22}.

\subsection{Low-field structure and temperature cuts}

The semiclassical magnetoviscous response is broad in field: increasing $|B|$ suppresses $\nu_{xx}^{(0)}$ through the Lorentz denominator, giving a positive magnetoconductivity or negative magnetoresistance in the (semiclassical) Gurzhi regime~\cite{Alekseev16,Alekseev20}.  The Cooperon correction adds a narrower scale.  At zero field, the coherent stress correction is maximal; at small finite field, the orbital Cooperon mass suppresses it and restores the stress relaxation rate toward its bare value.  The resistance can therefore show a small central dip near $B=0$, surrounded by nearby shoulder-like maxima, before the broader semiclassical negative magnetoresistance takes over.  A distinct feature is most visible when the zero-field stress correction is not parametrically small and when the Cooperon dephasing field lies below, or at least not far above, the semiclassical magnetoviscous field scale.

\begin{figure}[t]
    \centering
    \includegraphics[width=0.98\columnwidth]{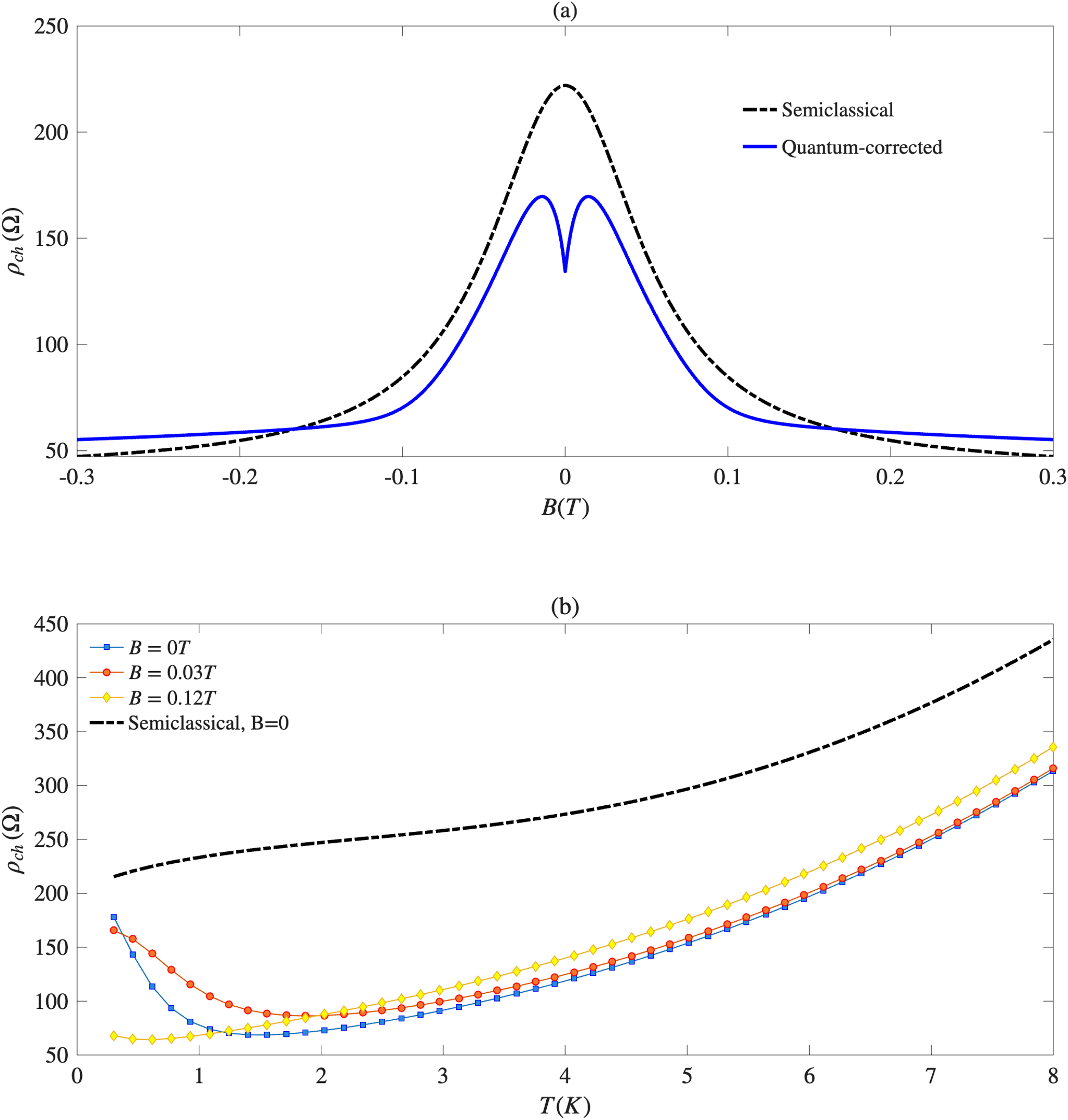}
    \caption{
    Representative magnetoviscous response including the hydrodynamic-Cooperon stress correction.
    (a) Channel magnetoresistance $\rho_{\rm ch}(B)$ at fixed temperature. The dashed black curve is the semiclassical result obtained from the bare stress relaxation rate $\Gamma_\pi(T)$, while the blue curve includes the Cooperon-induced correction $\Gamma_\pi\to\Gamma_\pi+\delta\Gamma_2(T,B)$. The quantum-corrected curve (solid blue line) develops a narrow low-field dip because the Cooperon correction is largest at $B=0$ and is suppressed by orbital magnetic dephasing at small finite field, while the broader background is controlled by the semiclassical magnetoviscous response (black dashed--dotted curve).
    (b) Fixed-field temperature cuts of the channel resistance. The low-field quantum-corrected curves show a nonmonotonic low-temperature dependence, absent in the semiclassical $B=0$ reference (black dashed--dotted line), reflecting the competition between the coherence-sensitive stress correction and the bare hydrodynamic background.    
    }
    \label{fig:quantum_magnetoviscosity}
\end{figure}

Figure~\ref{fig:quantum_magnetoviscosity} illustrates this competition.  Panel (a) separates the broad semiclassical magnetoviscous background from the narrower Cooperon-induced low-field structure.  The dashed--dotted curve is governed by the Lorentz suppression of the bare longitudinal viscosity, while the quantum-corrected curve contains an additional field scale set by the orbital dephasing of the hydrodynamic Cooperon.  Panel (b) shows the same physics in fixed-field temperature cuts: the correction is largest only in the window where coherence survives but the hydrodynamic stress mode is still active.  These cuts are especially useful because the Cooperon correction carries an explicit thermal fluctuation factor but is also reduced by stress relaxation, momentum relaxation, the infrared cutoff, and dephasing.  The resulting low-temperature profile need not be monotonic, and its detailed line shape is nonuniversal. Such nonmonotonic behavior is consistent with experimental observations\cite{Shi14}.

\subsection{Hall-viscous response}

The same stress-sector renormalization affects the transverse, or Hall-viscous, component of the magnetoviscosity tensor.  The symmetry is unchanged: $\nu_{xy}^{\rm eff}$ remains odd in $B$ and vanishes at $B=0$.  The correction instead suppresses the low-field Hall-viscosity slope through the replacement $\Gamma_\pi\to\Gamma_\pi+\delta\Gamma_2(T,B)$.  Since the orbital Cooperon mass is nonanalytic in weak field through the $|B|$ cutoff scale, this suppression can generate a sharp low-field structure in the Hall-viscous tensor.  In realistic devices, residual dephasing, finite-size cutoffs, inhomogeneity, and the nonlocal extraction protocol will round this structure~\cite{Avron98,Scaffidi17,Pellegrino17,Delacretaz17,Berdyugin19}.

\begin{figure}[t]
    \centering
    \includegraphics[width=\columnwidth]{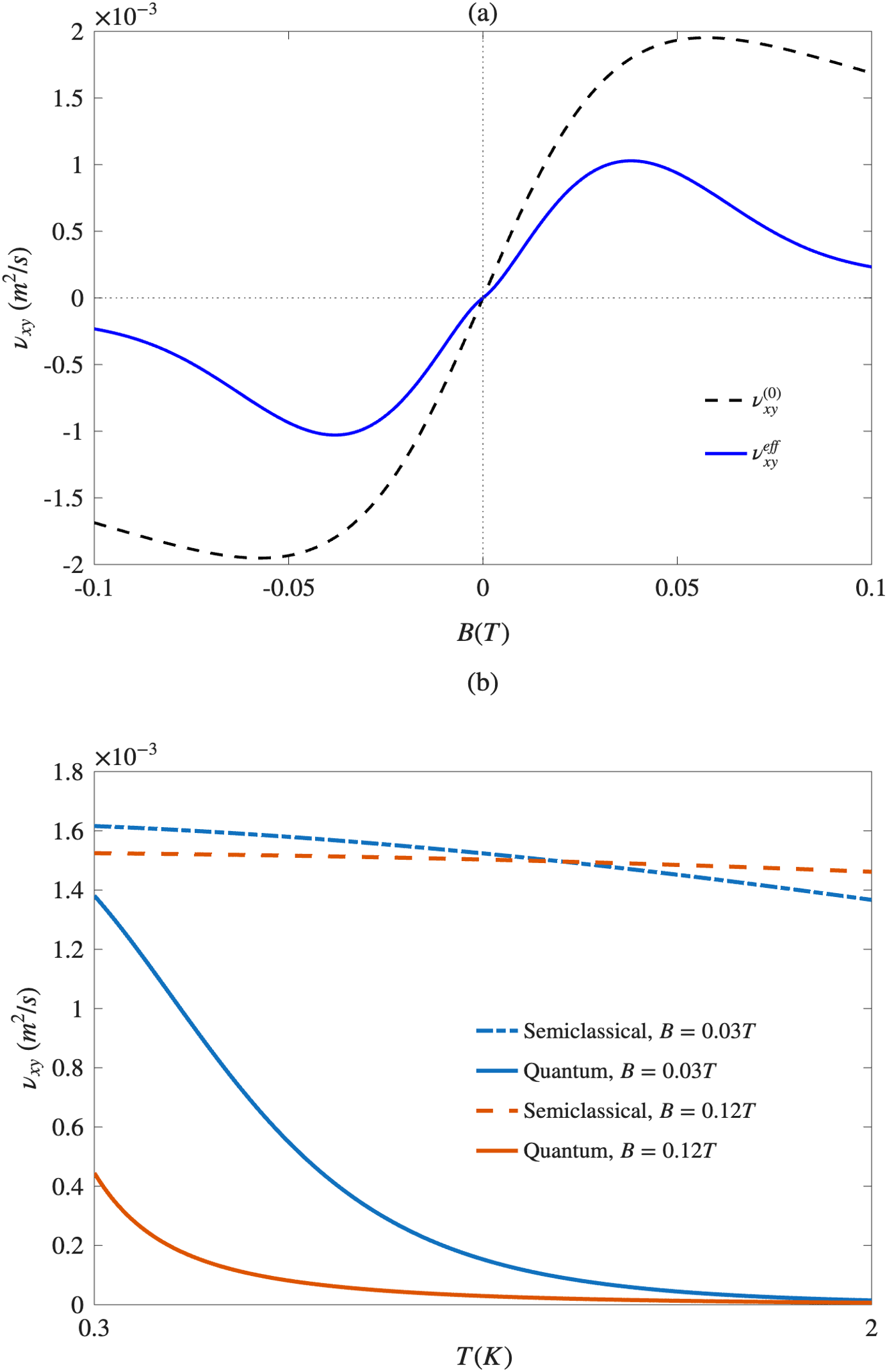}
    \caption{
    Quantum correction to the Hall-viscous component of the magnetoviscosity tensor.
    (a) Semiclassical (black dashed curve) and quantum-corrected (blue solid curve) Hall viscosities as a function of magnetic field.
    The Cooperon correction preserves the odd-in-$B$ symmetry of $\nu_{xy}$, but suppresses the low-field slope through
$\Gamma_\pi\to\Gamma_\pi+\delta\Gamma_2(B,T)$.
    (b) Temperature dependence of $\nu_{xy}$ at two representative values of the magnetic field, blue $B=0.03T$ and orange $B=0.12T$. In both cases the strong reduction of the quantum-corrected Hall viscosity with increasing temperature reflects the thermal dephasing effect of the hydrodynamic Cooperon in presence of magnetic fields.
    }
    \label{fig:hall_viscosity_quantum_corrected}
\end{figure}

Figure~\ref{fig:hall_viscosity_quantum_corrected},a shows this low--field kink structure. The Hall-viscosity kink is sharper than the corresponding magnetoresistive shoulder because it probes the reactive component of the viscosity tensor directly.  This makes it theoretically clean but experimentally fragile: temperature and dephasing can smooth the Hall-viscosity structure even when the longitudinal magnetoresistance retains a broader remnant of the same stress correction, as it can be appreciated in Fig.~\ref{fig:hall_viscosity_quantum_corrected},a, where a drastic reduction of the quantum correction to the Hall viscosity with temperature is shown.


\subsection{Comparison with existing experimental data}

High-mobility GaAs/AlGaAs heterostructures provide one relevant point of comparison.  Large negative magnetoresistance and narrow low-field structures have been reported in ultraclean two-dimensional electron gases, including regimes where purely quasiclassical memory-effect interpretations are incomplete~\cite{Renard08,Hatke12,Shi14,Gusev18}.  In particular, Ref.~\cite{Shi14} reports a broad negative-magnetoresistance background together with a much narrower central feature and low-temperature fixed-field cuts that are nonmonotonic or shoulder-like.  A more direct comparison to hydrodynamic magnetotransport is provided by the measurements of Gusev \emph{et al.}~\cite{Gusev18}, where viscous transport and Hall-viscosity signatures were reported in a mesoscopic two-dimensional electron system.  Their longitudinal magnetoresistivity displays low-field shoulder-like structures on top of the broader negative-magnetoresistance background, which is phenomenologically close to the response discussed here.

At the same time, such shoulders should not be regarded as unique fingerprints of a coherence-sensitive hydrodynamic Cooperon.  Raichev \emph{et al.}~\cite{Raichev20} showed that narrow two-dimensional conductors can display similar low-field magnetoresistance structures from a classical kinetic mechanism involving the size effect, partly diffusive boundary scattering, cyclotron-orbit commensurability, and electronic viscosity.  In that interpretation, the relevant field scales are set by the relation between the cyclotron diameter and the channel width, while increasing temperature enhances electron-electron scattering and progressively smears the ballistic commensurability structure.  This provides an important competing background against which any hydrodynamic-Cooperon contribution must be separated.

The temperature dependence may therefore provide a useful, although not definitive, discriminant between the two mechanisms.  In the classical size-effect scenario, the low-field shoulder is expected to be mostly suppressed as temperature drives the system away from the ballistic boundary-dominated regime and toward viscous flow.  In the present mechanism, the shoulder is instead a coherence-sensitive hydrodynamic-interference feature: it requires a developed hydrodynamic stress sector, but is suppressed once dephasing and the magnetic Cooperon mass become too large.  The resulting low-field magnetoresistive structure can therefore be strongest only in an intermediate window where hydrodynamic stress dynamics and phase coherence coexist.  This naturally allows a nonmonotonic temperature dependence, as illustrated in Fig.~\ref{fig:quantum_magnetoviscosity}, rather than a simple monotonic thermal smearing.

Our calculation is therefore not a quantitative fit to the GaAs data.  It isolates a different hydrodynamic mechanism by which magnetic orbital dephasing suppresses a Cooperon correction to the stress relaxation rate, thereby changing $\nu_{xx}^{\rm eff}(T,B)$ and the channel resistance.  The appropriate conclusion is qualitative: a coherence-sensitive stress-sector renormalization can naturally generate a narrow low-field structure in $\rho(B)$ and nonmonotonic low-temperature behavior in small fixed-field $\rho(T)$ cuts, while the detailed line shape remains sample- and geometry-dependent.  The comparison with Refs.~\cite{Gusev18,Raichev20} is useful precisely because it shows both sides of the issue: similar shoulders occur in experimental hydrodynamic magnetotransport, but classical size-effect and boundary mechanisms can also produce them.

Graphene Hall-viscosity measurements provide a second, more direct comparison to the magnetoviscous tensor. Berdyugin \emph{et al.} extracted the Hall viscosity of graphene's electron fluid from antisymmetric local-resistance measurements in the hydrodynamic regime, building on earlier graphene hydrodynamic signatures such as viscous backflow and superballistic flow~\cite{Bandurin16,KrishnaKumar17,Berdyugin19}.  Their main-text Hall-viscosity data are approximately linear in weak magnetic field and do not display a visible low-field kink of the type shown in Fig.~\ref{fig:hall_viscosity_quantum_corrected}.  This absence is not necessarily in conflict with the present mechanism.  In our calculation the kink in $\nu_{xy}^{\rm eff}$ is the most fragile tensor-level signature because it is controlled by dephasing, orbital cutoff scales, disorder averaging, finite-size effects, and the nonlocal inversion used to extract the Hall viscosity.

The longitudinal magnetoresistance in the same graphene experiment is more suggestive.  In Fig.~S5(a) of the Supplemental Material of Ref.~\cite{Berdyugin19}, the measured monolayer-graphene magnetoresistance in the hydrodynamic regime shows, at the lowest temperature displayed in that panel, a residual low-field shoulder-like structure around the central magnetoresistance feature.  The structure becomes less visible at higher temperatures.  The published hydrodynamic curves reproduce the broad negative magnetoresistance using the experimentally determined zero-field viscosity and no fitting parameters, but they do not isolate the residual low-field structure.  We therefore do not interpret those data as evidence for a hydrodynamic Cooperon.  Rather, they show that hydrodynamic longitudinal magnetoresistance can contain low-field structure in a temperature window where the extracted Hall viscosity itself appears smooth.  This hierarchy is compatible with our prediction: the Hall-viscosity kink is the more fragile signature, while the magnetoresistive shoulder can survive as a broader remnant of the same coherence-sensitive viscous correction.



\section{Discussion and outlook}
\label{sec:section_11}

The main result is a conservation-law projection of quantum interference in an electron fluid.  A uniform Cooperon correction cannot generate a dissipative mass for the conserved density or momentum sectors, so the first admissible dc correction appears in the nonconserved spin-two stress sector.  At zero field this gives a positive correction to the stress relaxation rate, lowers the shear viscosity, and lowers the viscous part of the Gurzhi channel resistance.  The sign is therefore opposite to ordinary weak localization because the interference mode renormalizes a hydrodynamic transport coefficient rather than the charge conductivity directly~\cite{Lee85,Altshuler82,Hershfield86,Bergmann84,Lucas18,Narozhny17,Fritz24}.

The SK random-friction construction makes this statement explicit.  Static random friction first produces a weak transverse momentum-relaxation rate in SCBA.  The same disorder vertex then generates a maximally crossed ladder of transverse hydrodynamic propagators.  Its infrared pole controls the stress self-energy through the pole mass, the ratio $D_C/Z_C$, and the spin-two vertex factor $Q^2/4$.  Cutoff-dependent analytic pieces renormalize local stress coefficients, while the sector, sign, and dephasing sensitivity follow from the low-energy pole and the Ward identities~\cite{Kamenev23,CrossleyGloriosoLiu17,GloriosoLiu18,Vollhardt80b,Wu22}.

The perpendicular-field extension is the independent result developed in the second part of the paper.  The field first changes the clean hydrodynamic fixed point, producing longitudinal and Hall viscosities through spin-one and spin-two cyclotron precession.  The cyclotron shift dresses the SCBA and Cooperon coefficients, but it does not by itself gap the interference mode.  The genuine magnetic dephasing scale is orbital: gauge covariance of the Cooperon center-of-mass coordinate produces a field-dependent contribution to the Cooperon mass proportional to $|B|D_C/Z_C$, which continuously returns to the zero-field Cooperon as $B\to0$~\cite{Lee85,Bergmann84,Altshuler82,Hershfield86,Alekseev16,Alekseev20,Avron98,Pellegrino17,Delacretaz17}.

In channel transport, the observable response is a competition between orbital dephasing and ordinary magnetoviscosity.  At weak field, dephasing suppresses the positive stress-relaxation correction and tends to restore the bare longitudinal viscosity.  At larger field, the semiclassical Lorentz denominator suppresses $\nu_{xx}$ directly.  The same replacement $\Gamma_\pi\to\Gamma_\pi+\delta\Gamma_2(T,B)$ also suppresses the low-field Hall-viscosity slope without generating a finite $\nu_{xy}$ at $B=0$.  The most favorable experimental regime is therefore a narrow Gurzhi channel with strong momentum-conserving relaxation, weak extrinsic momentum relaxation, long phase coherence, and a magnetic scale where orbital dephasing is resolved before the broader semiclassical magnetoviscous background dominates~\cite{Gurzhi63,deJong95,Molenkamp94,Gurzhi95,Jaggi91,Alekseev16,Alekseev20,Berdyugin19}.

Several extensions remain open.  A full vector Bethe--Salpeter treatment would incorporate the compressible longitudinal sector and its coupling to sound.  More general disorder ensembles remain an open direction.  Within the random-friction construction studied here, the model-dependent information is contained in $\tau_{\mathrm{mr}}^{-1}$, the pole mass, and the cutoff, while the stress-sector projection follows from the Ward identities and from the scalar transverse Cooperon pole.  Strongly interacting or quantum-critical fluids should admit the same scalar, vector, and spin-two organization without relying on quasiparticle harmonics.  In that broader setting, the present calculation should be viewed as a controlled kinetic realization of the principle that quantum interference in electron hydrodynamics renormalizes transport coefficients, with magnetic field providing a direct diagnostic of the coherent stress-sector correction~\cite{HartnollLucasSachdev18,DavisonGouterauxHartnoll15,DamleSachdev97,Wu22}.

\begin{acknowledgments}
The author is grateful to Matteo Baggioli, Karl Landsteiner, and Francisco Dom\'inguez-Adame for helpful comments on previous versions of this manuscript. The author also acknowledges financial support from the Ministerio de Ciencia e Innovaci\'on through Grant PID2024-161156NB-I00 and the Severo Ochoa Centres of Excellence program through Grant CEX2024-001445-S.
\end{acknowledgments}

\section*{Data Availability}

The numerical data used to generate the figures in this work are not publicly available. They are available from the corresponding author upon reasonable request.

\appendix

\section{Expansion of the zero-field Cooperon pole}
\label{app:zero-field-pole}

This appendix records the algebra leading from the cutoff-regulated inverse Cooperon, Eq.~\eqref{eq:Cinv_exact}, to the pole coefficients used in Sec.~V.  It is enough to write
\begin{eqnarray}
C_\perp^{-1}(\Omega,Q)
&=&\Delta^{-1}-\frac{1}{8\pi\nu}
\left[\ln(\nu\Lambda^2+\Gamma)-\ln\Gamma\right],
\nonumber\\
\Gamma(\Omega,Q)
&=&\tau_{\mathrm{mr}}^{-1}+\frac{\nu Q^2}{4}-\frac{\ii\Omega}{2}.
\end{eqnarray}
The mass is $C_0^{-1}=C_\perp^{-1}(0,0)$.  Differentiating the same expression at $\Omega=Q=0$ gives
\begin{eqnarray}
D_C
&=&\left.\frac{\partial C_\perp^{-1}}{\partial Q^2}\right|_0
=\frac{1}{32\pi}
\left[
\frac{1}{\tau_{\mathrm{mr}}^{-1}}
-
\frac{1}{\tau_{\mathrm{mr}}^{-1}+\nu\Lambda^2}
\right],
\nonumber\\
Z_C
&=&\left.\ii\frac{\partial C_\perp^{-1}}{\partial\Omega}\right|_0
=\frac{1}{16\pi\nu}
\left[
\frac{1}{\tau_{\mathrm{mr}}^{-1}}
-
\frac{1}{\tau_{\mathrm{mr}}^{-1}+\nu\Lambda^2}
\right].
\end{eqnarray}
These coefficients reproduce Eqs.~\eqref{eq:DC_final} and \eqref{eq:ZC_final}, while $C_0^{-1}$ gives Eq.~\eqref{eq:C0inv}.  The manipulations are exact within the scalar transverse ladder and the chosen sharp hydrodynamic cutoff.  Changing the ultraviolet regulator can shift nonuniversal pieces of $C_0^{-1}$, but not the infrared pole structure or the ratio $D_C/Z_C=\nu/2$.

\section{Stress self-energy from the Cooperon pole}
\label{app:stress-self-energy}

This appendix gives the algebra behind Eqs.~\eqref{eq:intCK_final} and \eqref{eq:deltaGamma_closed}.  Starting from the pole form \eqref{eq:CR_low}, the classical KMS limit gives
\begin{eqnarray}
C_\perp^K(\Omega,Q)
&\simeq&
\frac{4\ii T Z_C}{(A_\phi+D_CQ^2)^2+(Z_C\Omega)^2},
\nonumber\\
\int\frac{\dd\Omega}{2\pi}C_\perp^K(\Omega,Q)
&=&
\frac{2\ii T}{A_\phi+D_CQ^2}.
\end{eqnarray}
Substitution into the retarded stress self-energy then yields
\begin{eqnarray}
\Sigma_2^R(0,0)
&=&-
\frac{T}{4}\int\frac{\dd^2Q}{(2\pi)^2}
\frac{Q^2}{A_\phi+D_CQ^2},
\nonumber\\
\delta\Gamma_2(T)
&=&
\frac{T}{4}\int_{|\bm Q|<\Lambda_C}
\frac{\dd^2Q}{(2\pi)^2}
\frac{Q^2}{A_\phi+D_CQ^2}.
\end{eqnarray}
The radial integral gives the closed expression quoted in the main text,
\begin{eqnarray}
&\frac{T}{8\pi}&\int_0^{\Lambda_C}\dd Q\,
\frac{Q^3}{A_\phi+D_CQ^2}
=
\frac{T\Lambda_C^2}{16\pi D_C}
\nonumber\\
&-&
\frac{T A_\phi}{16\pi D_C^2}
\ln\!\left(1+\frac{D_C\Lambda_C^2}{A_\phi}\right).
\end{eqnarray}

\section{Magnetic harmonic reduction}
\label{app:mag-harmonic-reduction}

This appendix gives the short harmonic reduction underlying Sec.~VIII.  With the Fourier convention used in the main text, $\partial_t\to-\ii\omega$ and $\partial_\pm\to\ii q_\pm$, the Navier--Stokes truncation of Eq.~\eqref{eq:mag_harmonic_general} gives
\begin{eqnarray}
-\ii\omega a_0+\frac{\ii v_F}{2}(q_-a_{-1}+q_+a_1)&=&0,
\nonumber\\
(-\ii\omega-\ii\omega_c)a_1+\frac{\ii v_F}{2}(q_-a_0+q_+a_2)&=&e v_F E_+,
\nonumber\\
(-\ii\omega+\ii\omega_c)a_{-1}+\frac{\ii v_F}{2}(q_+a_0+q_-a_{-2})&=&e v_F E_-,
\nonumber\\
(-\ii\omega-2\ii\omega_c+\tau_{\mathrm{mc}}^{-1})a_2+\frac{\ii v_F}{2}q_-a_1&=&0,
\nonumber\\
(-\ii\omega+2\ii\omega_c+\tau_{\mathrm{mc}}^{-1})a_{-2}+\frac{\ii v_F}{2}q_+a_{-1}&=&0,
\end{eqnarray}
where $E_\pm=E_x\pm\ii E_y$.  The last two equations give
\begin{eqnarray}
a_2
&=&-
\frac{\ii v_F}{2}
\frac{q_-}{\tau_{\mathrm{mc}}^{-1}(T)-2\ii\omega_c-\ii\omega}
\,a_1,
\nonumber\\
a_{-2}
&=&-
\frac{\ii v_F}{2}
\frac{q_+}{\tau_{\mathrm{mc}}^{-1}(T)+2\ii\omega_c-\ii\omega}
\,a_{-1}.
\end{eqnarray}
At Navier--Stokes order one may set $|\omega|\tau_{\mathrm{mc}}\ll1$ in these denominators.  Substitution into the $m=\pm1$ equations produces the complex coefficient \eqref{eq:complex_bare_mag_visc}.  Finally, using $u_x=v_F(a_1+a_{-1})/2$ and $u_y=\ii v_F(a_1-a_{-1})/2$ gives Eq.~\eqref{eq:clean_mag_hydro_eq}.  This derivation is identical in structure to the zero-field viscosity reduction, except that the stress eigenvalues are shifted by $\mp2\ii\omega_c$.

\section{Magnetic Cooperon pole coefficients}
\label{app:mag-cooperon-coeffs}

This appendix records the coefficient formulas obtained by expanding Eq.~\eqref{eq:mag_bubble_reduced}.  To keep the expressions compact, define $\gamma_B=\tau_{\mathrm{mr}}^{-1}(T,B)$, $\nu_B=\nu_{xx}^{(0)}(T,B)$, $\nu_H=\nu_{xy}^{(0)}(T,B)$, and $E_q=\gamma_B+\nu_B q^2$.  The static bubble, frequency coefficient, and stiffness are
\begin{eqnarray}
\Pi_B(0,0;T,B)
&=&\frac12\int\frac{\dd^2q}{(2\pi)^2}\frac{1}{E_q},
\nonumber\\
Z_C(T,B)
&=&\frac14\int\frac{\dd^2q}{(2\pi)^2}\frac{1}{E_q^2},
\nonumber\\
D_C(T,B)
&=&\int\frac{\dd^2q}{(2\pi)^2}
\left[
\frac{\nu_B}{8E_q^2}
+
\frac{\nu_H^2q^2}{4E_q^3}
\right].
\end{eqnarray}
The second contribution to $D_C$ comes from the square of the Hall-viscous reactive shift in Eq.~\eqref{eq:mag_bubble_reduced}; it survives angular averaging because it is even in $B$.

For a sharp hydrodynamic cutoff $\Lambda$, the same coefficients become
\begin{eqnarray}
\Pi_B(0,0;T,B)
&=&\frac{1}{8\pi\nu_B}
\ln\left(1+\frac{\nu_B\Lambda^2}{\gamma_B}\right),
\nonumber\\
Z_C(T,B)
&=&\frac{1}{16\pi\nu_B}
\left[
\frac{1}{\gamma_B}
-
\frac{1}{\gamma_B+\nu_B\Lambda^2}
\right],
\nonumber\\
D_C(T,B)
&=&\frac{1}{32\pi}
\left[
\frac{1}{\gamma_B}
-
\frac{1}{\gamma_B+\nu_B\Lambda^2}
\right]
\nonumber\\
&&+
\frac{\nu_H^2\Lambda^4}{32\pi\gamma_B(\gamma_B+\nu_B\Lambda^2)^2}.
\end{eqnarray}
In the weak-damping limit these expressions give Eq.~\eqref{eq:mag_Dhyd_weak}.  Regulator-dependent pieces can modify the ultraviolet part of $C_0^{-1}$, but the Hall-viscous contribution to the pole stiffness follows from the infrared expansion of the retarded--advanced ladder.


\begin{thebibliography}{36}%
\makeatletter
\providecommand \@ifxundefined [1]{%
 \@ifx{#1\undefined}
}%
\providecommand \@ifnum [1]{%
 \ifnum #1\expandafter \@firstoftwo
 \else \expandafter \@secondoftwo
 \fi
}%
\providecommand \@ifx [1]{%
 \ifx #1\expandafter \@firstoftwo
 \else \expandafter \@secondoftwo
 \fi
}%
\providecommand \natexlab [1]{#1}%
\providecommand \enquote  [1]{``#1''}%
\providecommand \bibnamefont  [1]{#1}%
\providecommand \bibfnamefont [1]{#1}%
\providecommand \citenamefont [1]{#1}%
\providecommand \href@noop [0]{\@secondoftwo}%
\providecommand \href [0]{\begingroup \@sanitize@url \@href}%
\providecommand \@href[1]{\@@startlink{#1}\@@href}%
\providecommand \@@href[1]{\endgroup#1\@@endlink}%
\providecommand \@sanitize@url [0]{\catcode `\\12\catcode `\$12\catcode
  `\&12\catcode `\#12\catcode `\^12\catcode `\_12\catcode `\%12\relax}%
\providecommand \@@startlink[1]{}%
\providecommand \@@endlink[0]{}%
\providecommand \url  [0]{\begingroup\@sanitize@url \@url }%
\providecommand \@url [1]{\endgroup\@href {#1}{\urlprefix }}%
\providecommand \urlprefix  [0]{URL }%
\providecommand \Eprint [0]{\href }%
\providecommand \doibase [0]{https://doi.org/}%
\providecommand \selectlanguage [0]{\@gobble}%
\providecommand \bibinfo  [0]{\@secondoftwo}%
\providecommand \bibfield  [0]{\@secondoftwo}%
\providecommand \translation [1]{[#1]}%
\providecommand \BibitemOpen [0]{}%
\providecommand \bibitemStop [0]{}%
\providecommand \bibitemNoStop [0]{.\EOS\space}%
\providecommand \EOS [0]{\spacefactor3000\relax}%
\providecommand \BibitemShut  [1]{\csname bibitem#1\endcsname}%
\let\auto@bib@innerbib\@empty
\bibitem [{\citenamefont {Lee}\ and\ \citenamefont
  {Ramakrishnan}(1985)}]{Lee85}%
  \BibitemOpen
  \bibfield  {author} {\bibinfo {author} {\bibfnamefont {P.~A.}\ \bibnamefont
  {Lee}}\ and\ \bibinfo {author} {\bibfnamefont {T.~V.}\ \bibnamefont
  {Ramakrishnan}},\ }\bibfield  {title} {\bibinfo {title} {Disordered
  electronic systems},\ }\href {https://doi.org/10.1103/RevModPhys.57.287}
  {\bibfield  {journal} {\bibinfo  {journal} {Rev. Mod. Phys.}\ }\textbf
  {\bibinfo {volume} {57}},\ \bibinfo {pages} {287} (\bibinfo {year}
  {1985})}\BibitemShut {NoStop}%
\bibitem [{\citenamefont {Altshuler}\ \emph {et~al.}(1982)\citenamefont
  {Altshuler}, \citenamefont {Aronov},\ and\ \citenamefont
  {Khmelnitskii}}]{Altshuler82}%
  \BibitemOpen
  \bibfield  {author} {\bibinfo {author} {\bibfnamefont {B.~L.}\ \bibnamefont
  {Altshuler}}, \bibinfo {author} {\bibfnamefont {A.~G.}\ \bibnamefont
  {Aronov}},\ and\ \bibinfo {author} {\bibfnamefont {D.~E.}\ \bibnamefont
  {Khmelnitskii}},\ }\bibfield  {title} {\bibinfo {title} {Effects of
  electron-electron collisions with small energy transfers on quantum
  localisation},\ }\href {https://doi.org/10.1088/0022-3719/15/36/018}
  {\bibfield  {journal} {\bibinfo  {journal} {Journal of Physics C: Solid State
  Physics}\ }\textbf {\bibinfo {volume} {15}},\ \bibinfo {pages} {7367}
  (\bibinfo {year} {1982})}\BibitemShut {NoStop}%
\bibitem [{\citenamefont {Hershfield}\ and\ \citenamefont
  {Ambegaokar}(1986)}]{Hershfield86}%
  \BibitemOpen
  \bibfield  {author} {\bibinfo {author} {\bibfnamefont {S.}~\bibnamefont
  {Hershfield}}\ and\ \bibinfo {author} {\bibfnamefont {V.}~\bibnamefont
  {Ambegaokar}},\ }\bibfield  {title} {\bibinfo {title} {Transport equation for
  weakly localized electrons},\ }\href
  {https://doi.org/10.1103/PhysRevB.34.2147} {\bibfield  {journal} {\bibinfo
  {journal} {Phys. Rev. B}\ }\textbf {\bibinfo {volume} {34}},\ \bibinfo
  {pages} {2147} (\bibinfo {year} {1986})}\BibitemShut {NoStop}%
\bibitem [{\citenamefont {Bergmann}(1984)}]{Bergmann84}%
  \BibitemOpen
  \bibfield  {author} {\bibinfo {author} {\bibfnamefont {G.}~\bibnamefont
  {Bergmann}},\ }\bibfield  {title} {\bibinfo {title} {Weak localization in
  thin films: a time-of-flight experiment with conduction electrons},\ }\href
  {https://doi.org/https://doi.org/10.1016/0370-1573(84)90103-0} {\bibfield
  {journal} {\bibinfo  {journal} {Physics Reports}\ }\textbf {\bibinfo {volume}
  {107}},\ \bibinfo {pages} {1} (\bibinfo {year} {1984})}\BibitemShut {NoStop}%
\bibitem [{\citenamefont {Gurzhi}(1963)}]{Gurzhi63}%
  \BibitemOpen
  \bibfield  {author} {\bibinfo {author} {\bibfnamefont {R.~N.}\ \bibnamefont
  {Gurzhi}},\ }\bibfield  {title} {\bibinfo {title} {Minimum of resistance in
  impurity-free conductors},\ }\href@noop {} {\bibfield  {journal} {\bibinfo
  {journal} {Soviet Physics JETP}\ }\textbf {\bibinfo {volume} {17}},\ \bibinfo
  {pages} {521} (\bibinfo {year} {1963})}\BibitemShut {NoStop}%
\bibitem [{\citenamefont {de~Jong}\ and\ \citenamefont
  {Molenkamp}(1995)}]{deJong95}%
  \BibitemOpen
  \bibfield  {author} {\bibinfo {author} {\bibfnamefont {M.~J.~M.}\
  \bibnamefont {de~Jong}}\ and\ \bibinfo {author} {\bibfnamefont {L.~W.}\
  \bibnamefont {Molenkamp}},\ }\bibfield  {title} {\bibinfo {title}
  {Hydrodynamic electron flow in high-mobility wires},\ }\href
  {https://doi.org/10.1103/PhysRevB.51.13389} {\bibfield  {journal} {\bibinfo
  {journal} {Physical Review B}\ }\textbf {\bibinfo {volume} {51}},\ \bibinfo
  {pages} {13389} (\bibinfo {year} {1995})}\BibitemShut {NoStop}%
\bibitem [{\citenamefont {Molenkamp}\ and\ \citenamefont
  {de~Jong}(1994)}]{Molenkamp94}%
  \BibitemOpen
  \bibfield  {author} {\bibinfo {author} {\bibfnamefont {L.~W.}\ \bibnamefont
  {Molenkamp}}\ and\ \bibinfo {author} {\bibfnamefont {M.~J.~M.}\ \bibnamefont
  {de~Jong}},\ }\bibfield  {title} {\bibinfo {title} {Observation of knudsen
  and gurzhi transport regimes in a two-dimensional wire},\ }\href
  {https://doi.org/10.1016/0038-1101(94)90244-5} {\bibfield  {journal}
  {\bibinfo  {journal} {Solid-State Electronics}\ }\textbf {\bibinfo {volume}
  {37}},\ \bibinfo {pages} {551} (\bibinfo {year} {1994})}\BibitemShut
  {NoStop}%
\bibitem [{\citenamefont {Gurzhi}\ \emph {et~al.}(1995)\citenamefont {Gurzhi},
  \citenamefont {Kalinenko},\ and\ \citenamefont {Kopeliovich}}]{Gurzhi95}%
  \BibitemOpen
  \bibfield  {author} {\bibinfo {author} {\bibfnamefont {R.~N.}\ \bibnamefont
  {Gurzhi}}, \bibinfo {author} {\bibfnamefont {A.~N.}\ \bibnamefont
  {Kalinenko}},\ and\ \bibinfo {author} {\bibfnamefont {A.~I.}\ \bibnamefont
  {Kopeliovich}},\ }\bibfield  {title} {\bibinfo {title} {Electron-electron
  collisions and a new hydrodynamic effect in two-dimensional electron gas},\
  }\href {https://doi.org/10.1103/PhysRevLett.74.3872} {\bibfield  {journal}
  {\bibinfo  {journal} {Physical Review Letters}\ }\textbf {\bibinfo {volume}
  {74}},\ \bibinfo {pages} {3872} (\bibinfo {year} {1995})}\BibitemShut
  {NoStop}%
\bibitem [{\citenamefont {Bandurin}\ \emph {et~al.}(2016)\citenamefont
  {Bandurin}, \citenamefont {Torre}, \citenamefont {Krishna~Kumar},
  \citenamefont {Ben~Shalom}, \citenamefont {Tomadin}, \citenamefont
  {Principi}, \citenamefont {Auton}, \citenamefont {Khestanova}, \citenamefont
  {Novoselov}, \citenamefont {Grigorieva}, \citenamefont {Ponomarenko},
  \citenamefont {Geim},\ and\ \citenamefont {Polini}}]{Bandurin16}%
  \BibitemOpen
  \bibfield  {author} {\bibinfo {author} {\bibfnamefont {D.~A.}\ \bibnamefont
  {Bandurin}}, \bibinfo {author} {\bibfnamefont {I.}~\bibnamefont {Torre}},
  \bibinfo {author} {\bibfnamefont {R.}~\bibnamefont {Krishna~Kumar}}, \bibinfo
  {author} {\bibfnamefont {M.}~\bibnamefont {Ben~Shalom}}, \bibinfo {author}
  {\bibfnamefont {A.}~\bibnamefont {Tomadin}}, \bibinfo {author} {\bibfnamefont
  {A.}~\bibnamefont {Principi}}, \bibinfo {author} {\bibfnamefont {G.~H.}\
  \bibnamefont {Auton}}, \bibinfo {author} {\bibfnamefont {E.}~\bibnamefont
  {Khestanova}}, \bibinfo {author} {\bibfnamefont {K.~S.}\ \bibnamefont
  {Novoselov}}, \bibinfo {author} {\bibfnamefont {I.~V.}\ \bibnamefont
  {Grigorieva}}, \bibinfo {author} {\bibfnamefont {L.~A.}\ \bibnamefont
  {Ponomarenko}}, \bibinfo {author} {\bibfnamefont {A.~K.}\ \bibnamefont
  {Geim}},\ and\ \bibinfo {author} {\bibfnamefont {M.}~\bibnamefont {Polini}},\
  }\bibfield  {title} {\bibinfo {title} {Negative local resistance caused by
  viscous electron backflow in graphene},\ }\href
  {https://doi.org/10.1126/science.aad0201} {\bibfield  {journal} {\bibinfo
  {journal} {Science}\ }\textbf {\bibinfo {volume} {351}},\ \bibinfo {pages}
  {1055} (\bibinfo {year} {2016})}\BibitemShut {NoStop}%
\bibitem [{\citenamefont {Krishna~Kumar}\ \emph {et~al.}(2017)\citenamefont
  {Krishna~Kumar} \emph {et~al.}}]{KrishnaKumar17}%
  \BibitemOpen
  \bibfield  {author} {\bibinfo {author} {\bibfnamefont {R.}~\bibnamefont
  {Krishna~Kumar}} \emph {et~al.},\ }\bibfield  {title} {\bibinfo {title}
  {Superballistic flow of viscous electron fluid through graphene
  constrictions},\ }\href {https://doi.org/10.1038/nphys4240} {\bibfield
  {journal} {\bibinfo  {journal} {Nature Physics}\ }\textbf {\bibinfo {volume}
  {13}},\ \bibinfo {pages} {1182} (\bibinfo {year} {2017})}\BibitemShut
  {NoStop}%
\bibitem [{\citenamefont {Berdyugin}\ \emph {et~al.}(2019)\citenamefont
  {Berdyugin}, \citenamefont {Xu}, \citenamefont {Pellegrino}, \citenamefont
  {Kumar}, \citenamefont {Principi}, \citenamefont {Torre}, \citenamefont
  {Shalom}, \citenamefont {Taniguchi}, \citenamefont {Watanabe}, \citenamefont
  {Grigorieva}, \citenamefont {Polini}, \citenamefont {Geim},\ and\
  \citenamefont {Bandurin}}]{Berdyugin19}%
  \BibitemOpen
  \bibfield  {author} {\bibinfo {author} {\bibfnamefont {A.~I.}\ \bibnamefont
  {Berdyugin}}, \bibinfo {author} {\bibfnamefont {S.~G.}\ \bibnamefont {Xu}},
  \bibinfo {author} {\bibfnamefont {F.~M.~D.}\ \bibnamefont {Pellegrino}},
  \bibinfo {author} {\bibfnamefont {R.~K.}\ \bibnamefont {Kumar}}, \bibinfo
  {author} {\bibfnamefont {A.}~\bibnamefont {Principi}}, \bibinfo {author}
  {\bibfnamefont {I.}~\bibnamefont {Torre}}, \bibinfo {author} {\bibfnamefont
  {M.~B.}\ \bibnamefont {Shalom}}, \bibinfo {author} {\bibfnamefont
  {T.}~\bibnamefont {Taniguchi}}, \bibinfo {author} {\bibfnamefont
  {K.}~\bibnamefont {Watanabe}}, \bibinfo {author} {\bibfnamefont {I.~V.}\
  \bibnamefont {Grigorieva}}, \bibinfo {author} {\bibfnamefont
  {M.}~\bibnamefont {Polini}}, \bibinfo {author} {\bibfnamefont {A.~K.}\
  \bibnamefont {Geim}},\ and\ \bibinfo {author} {\bibfnamefont {D.~A.}\
  \bibnamefont {Bandurin}},\ }\bibfield  {title} {\bibinfo {title} {Measuring
  hall viscosity of graphene’s electron fluid},\ }\href
  {https://doi.org/10.1126/science.aau0685} {\bibfield  {journal} {\bibinfo
  {journal} {Science}\ }\textbf {\bibinfo {volume} {364}},\ \bibinfo {pages}
  {162} (\bibinfo {year} {2019})},\ \Eprint
  {https://arxiv.org/abs/https://www.science.org/doi/pdf/10.1126/science.aau0685}
  {https://www.science.org/doi/pdf/10.1126/science.aau0685} \BibitemShut
  {NoStop}%
\bibitem [{\citenamefont {Lucas}\ and\ \citenamefont {Fong}(2018)}]{Lucas18}%
  \BibitemOpen
  \bibfield  {author} {\bibinfo {author} {\bibfnamefont {A.}~\bibnamefont
  {Lucas}}\ and\ \bibinfo {author} {\bibfnamefont {K.~C.}\ \bibnamefont
  {Fong}},\ }\bibfield  {title} {\bibinfo {title} {Hydrodynamics of electrons
  in graphene},\ }\href {https://doi.org/10.1088/1361-648X/aaa274} {\bibfield
  {journal} {\bibinfo  {journal} {Journal of Physics: Condensed Matter}\
  }\textbf {\bibinfo {volume} {30}},\ \bibinfo {pages} {053001} (\bibinfo
  {year} {2018})}\BibitemShut {NoStop}%
\bibitem [{\citenamefont {Narozhny}\ \emph {et~al.}(2017)\citenamefont
  {Narozhny}, \citenamefont {Gornyi}, \citenamefont {Mirlin},\ and\
  \citenamefont {Schmalian}}]{Narozhny17}%
  \BibitemOpen
  \bibfield  {author} {\bibinfo {author} {\bibfnamefont {B.~N.}\ \bibnamefont
  {Narozhny}}, \bibinfo {author} {\bibfnamefont {I.~V.}\ \bibnamefont
  {Gornyi}}, \bibinfo {author} {\bibfnamefont {A.~D.}\ \bibnamefont {Mirlin}},\
  and\ \bibinfo {author} {\bibfnamefont {J.}~\bibnamefont {Schmalian}},\
  }\bibfield  {title} {\bibinfo {title} {Hydrodynamic approach to electronic
  transport in graphene},\ }\href {https://doi.org/10.1002/andp.201700043}
  {\bibfield  {journal} {\bibinfo  {journal} {Annalen der Physik}\ }\textbf
  {\bibinfo {volume} {529}},\ \bibinfo {pages} {1700043} (\bibinfo {year}
  {2017})}\BibitemShut {NoStop}%
\bibitem [{\citenamefont {Fritz}\ and\ \citenamefont
  {Scaffidi}(2024)}]{Fritz24}%
  \BibitemOpen
  \bibfield  {author} {\bibinfo {author} {\bibfnamefont {L.}~\bibnamefont
  {Fritz}}\ and\ \bibinfo {author} {\bibfnamefont {T.}~\bibnamefont
  {Scaffidi}},\ }\bibfield  {title} {\bibinfo {title} {Hydrodynamic electronic
  transport},\ }\href
  {https://doi.org/10.1146/annurev-conmatphys-040521-042014} {\bibfield
  {journal} {\bibinfo  {journal} {Annual Review of Condensed Matter Physics}\
  }\textbf {\bibinfo {volume} {15}},\ \bibinfo {pages} {17} (\bibinfo {year}
  {2024})}\BibitemShut {NoStop}%
\bibitem [{\citenamefont {Varnavides}\ \emph {et~al.}(2023)\citenamefont
  {Varnavides}, \citenamefont {Yacoby}, \citenamefont {Felser},\ and\
  \citenamefont {Narang}}]{Varnavides23}%
  \BibitemOpen
  \bibfield  {author} {\bibinfo {author} {\bibfnamefont {G.}~\bibnamefont
  {Varnavides}}, \bibinfo {author} {\bibfnamefont {A.}~\bibnamefont {Yacoby}},
  \bibinfo {author} {\bibfnamefont {C.}~\bibnamefont {Felser}},\ and\ \bibinfo
  {author} {\bibfnamefont {P.}~\bibnamefont {Narang}},\ }\bibfield  {title}
  {\bibinfo {title} {Charge transport and hydrodynamics in materials},\ }\href
  {https://doi.org/10.1038/s41578-023-00597-3} {\bibfield  {journal} {\bibinfo
  {journal} {Nature Reviews Materials}\ }\textbf {\bibinfo {volume} {8}},\
  \bibinfo {pages} {726} (\bibinfo {year} {2023})}\BibitemShut {NoStop}%
\bibitem [{\citenamefont {Crossley}\ \emph {et~al.}(2017)\citenamefont
  {Crossley}, \citenamefont {Glorioso},\ and\ \citenamefont
  {Liu}}]{CrossleyGloriosoLiu17}%
  \BibitemOpen
  \bibfield  {author} {\bibinfo {author} {\bibfnamefont {M.}~\bibnamefont
  {Crossley}}, \bibinfo {author} {\bibfnamefont {P.}~\bibnamefont {Glorioso}},\
  and\ \bibinfo {author} {\bibfnamefont {H.}~\bibnamefont {Liu}},\ }\bibfield
  {title} {\bibinfo {title} {Effective field theory of dissipative fluids},\
  }\href {https://doi.org/10.1007/JHEP09(2017)095} {\bibfield  {journal}
  {\bibinfo  {journal} {Journal of High Energy Physics}\ }\textbf {\bibinfo
  {volume} {2017}},\ \bibinfo {pages} {095} (\bibinfo {year} {2017})},\ \Eprint
  {https://arxiv.org/abs/1511.03646} {arXiv:1511.03646 [hep-th]} \BibitemShut
  {NoStop}%
\bibitem [{\citenamefont {Glorioso}\ and\ \citenamefont
  {Liu}(2018)}]{GloriosoLiu18}%
  \BibitemOpen
  \bibfield  {author} {\bibinfo {author} {\bibfnamefont {P.}~\bibnamefont
  {Glorioso}}\ and\ \bibinfo {author} {\bibfnamefont {H.}~\bibnamefont {Liu}},\
  }\bibfield  {title} {\bibinfo {title} {Lectures on non-equilibrium effective
  field theories and fluctuating hydrodynamics},\ }\href
  {https://doi.org/10.22323/1.305.0008} {\bibfield  {journal} {\bibinfo
  {journal} {Proceedings of Science}\ }\textbf {\bibinfo {volume} {TASI2017}},\
  \bibinfo {pages} {008} (\bibinfo {year} {2018})},\ \Eprint
  {https://arxiv.org/abs/1805.09331} {arXiv:1805.09331 [hep-th]} \BibitemShut
  {NoStop}%
\bibitem [{\citenamefont {Kamenev}(2023)}]{Kamenev23}%
  \BibitemOpen
  \bibfield  {author} {\bibinfo {author} {\bibfnamefont {A.}~\bibnamefont
  {Kamenev}},\ }\href {https://books.google.es/books?id=aDijEAAAQBAJ} {\emph
  {\bibinfo {title} {Field Theory of Non-Equilibrium Systems}}}\ (\bibinfo
  {publisher} {Cambridge University Press},\ \bibinfo {year}
  {2023})\BibitemShut {NoStop}%
\bibitem [{\citenamefont {Vollhardt}\ and\ \citenamefont
  {W\"olfle}(1980)}]{Vollhardt80b}%
  \BibitemOpen
  \bibfield  {author} {\bibinfo {author} {\bibfnamefont {D.}~\bibnamefont
  {Vollhardt}}\ and\ \bibinfo {author} {\bibfnamefont {P.}~\bibnamefont
  {W\"olfle}},\ }\bibfield  {title} {\bibinfo {title} {Diagrammatic,
  self-consistent treatment of the anderson localization problem in
  $d\ensuremath{\le}2$ dimensions},\ }\href
  {https://doi.org/10.1103/PhysRevB.22.4666} {\bibfield  {journal} {\bibinfo
  {journal} {Phys. Rev. B}\ }\textbf {\bibinfo {volume} {22}},\ \bibinfo
  {pages} {4666} (\bibinfo {year} {1980})}\BibitemShut {NoStop}%
\bibitem [{\citenamefont {Alekseev}(2016)}]{Alekseev16}%
  \BibitemOpen
  \bibfield  {author} {\bibinfo {author} {\bibfnamefont {P.~S.}\ \bibnamefont
  {Alekseev}},\ }\bibfield  {title} {\bibinfo {title} {Negative
  magnetoresistance in viscous flow of two-dimensional electrons},\ }\href
  {https://doi.org/10.1103/PhysRevLett.117.166601} {\bibfield  {journal}
  {\bibinfo  {journal} {Physical Review Letters}\ }\textbf {\bibinfo {volume}
  {117}},\ \bibinfo {pages} {166601} (\bibinfo {year} {2016})}\BibitemShut
  {NoStop}%
\bibitem [{\citenamefont {Alekseev}\ and\ \citenamefont
  {Dmitriev}(2020)}]{Alekseev20}%
  \BibitemOpen
  \bibfield  {author} {\bibinfo {author} {\bibfnamefont {P.~S.}\ \bibnamefont
  {Alekseev}}\ and\ \bibinfo {author} {\bibfnamefont {A.~P.}\ \bibnamefont
  {Dmitriev}},\ }\bibfield  {title} {\bibinfo {title} {Viscosity of
  two-dimensional electrons},\ }\href
  {https://doi.org/10.1103/PhysRevB.102.241409} {\bibfield  {journal} {\bibinfo
   {journal} {Physical Review B}\ }\textbf {\bibinfo {volume} {102}},\ \bibinfo
  {pages} {241409} (\bibinfo {year} {2020})}\BibitemShut {NoStop}%
\bibitem [{\citenamefont {Avron}(1998)}]{Avron98}%
  \BibitemOpen
  \bibfield  {author} {\bibinfo {author} {\bibfnamefont {J.~E.}\ \bibnamefont
  {Avron}},\ }\bibfield  {title} {\bibinfo {title} {Odd viscosity},\ }\href
  {https://doi.org/10.1023/A:1023084404080} {\bibfield  {journal} {\bibinfo
  {journal} {Journal of Statistical Physics}\ }\textbf {\bibinfo {volume}
  {92}},\ \bibinfo {pages} {543} (\bibinfo {year} {1998})}\BibitemShut
  {NoStop}%
\bibitem [{\citenamefont {Scaffidi}\ \emph {et~al.}(2017)\citenamefont
  {Scaffidi}, \citenamefont {Nandi}, \citenamefont {Schmidt}, \citenamefont
  {Mackenzie},\ and\ \citenamefont {Moore}}]{Scaffidi17}%
  \BibitemOpen
  \bibfield  {author} {\bibinfo {author} {\bibfnamefont {T.}~\bibnamefont
  {Scaffidi}}, \bibinfo {author} {\bibfnamefont {N.}~\bibnamefont {Nandi}},
  \bibinfo {author} {\bibfnamefont {B.}~\bibnamefont {Schmidt}}, \bibinfo
  {author} {\bibfnamefont {A.~P.}\ \bibnamefont {Mackenzie}},\ and\ \bibinfo
  {author} {\bibfnamefont {J.~E.}\ \bibnamefont {Moore}},\ }\bibfield  {title}
  {\bibinfo {title} {Hydrodynamic electron flow and hall viscosity},\ }\href
  {https://doi.org/10.1103/PhysRevLett.118.226601} {\bibfield  {journal}
  {\bibinfo  {journal} {Physical Review Letters}\ }\textbf {\bibinfo {volume}
  {118}},\ \bibinfo {pages} {226601} (\bibinfo {year} {2017})}\BibitemShut
  {NoStop}%
\bibitem [{\citenamefont {Pellegrino}\ \emph {et~al.}(2017)\citenamefont
  {Pellegrino}, \citenamefont {Torre},\ and\ \citenamefont
  {Polini}}]{Pellegrino17}%
  \BibitemOpen
  \bibfield  {author} {\bibinfo {author} {\bibfnamefont {F.~M.~D.}\
  \bibnamefont {Pellegrino}}, \bibinfo {author} {\bibfnamefont
  {I.}~\bibnamefont {Torre}},\ and\ \bibinfo {author} {\bibfnamefont
  {M.}~\bibnamefont {Polini}},\ }\bibfield  {title} {\bibinfo {title} {Nonlocal
  transport and the hall viscosity of two-dimensional hydrodynamic electron
  liquids},\ }\href {https://doi.org/10.1103/PhysRevB.96.195401} {\bibfield
  {journal} {\bibinfo  {journal} {Physical Review B}\ }\textbf {\bibinfo
  {volume} {96}},\ \bibinfo {pages} {195401} (\bibinfo {year}
  {2017})}\BibitemShut {NoStop}%
\bibitem [{\citenamefont {Delacr{\'e}taz}\ and\ \citenamefont
  {Gromov}(2017)}]{Delacretaz17}%
  \BibitemOpen
  \bibfield  {author} {\bibinfo {author} {\bibfnamefont {L.~V.}\ \bibnamefont
  {Delacr{\'e}taz}}\ and\ \bibinfo {author} {\bibfnamefont {A.}~\bibnamefont
  {Gromov}},\ }\bibfield  {title} {\bibinfo {title} {Transport signatures of
  the hall viscosity},\ }\href {https://doi.org/10.1103/PhysRevLett.119.226602}
  {\bibfield  {journal} {\bibinfo  {journal} {Physical Review Letters}\
  }\textbf {\bibinfo {volume} {119}},\ \bibinfo {pages} {226602} (\bibinfo
  {year} {2017})}\BibitemShut {NoStop}%
\bibitem [{\citenamefont {Kamenev}\ and\ \citenamefont
  {Andreev}(1999)}]{Kamenev99}%
  \BibitemOpen
  \bibfield  {author} {\bibinfo {author} {\bibfnamefont {A.}~\bibnamefont
  {Kamenev}}\ and\ \bibinfo {author} {\bibfnamefont {A.}~\bibnamefont
  {Andreev}},\ }\bibfield  {title} {\bibinfo {title} {Electron-electron
  interactions in disordered metals: Keldysh formalism},\ }\href
  {https://doi.org/10.1103/PhysRevB.60.2218} {\bibfield  {journal} {\bibinfo
  {journal} {Phys. Rev. B}\ }\textbf {\bibinfo {volume} {60}},\ \bibinfo
  {pages} {2218} (\bibinfo {year} {1999})}\BibitemShut {NoStop}%
\bibitem [{\citenamefont {Wu}\ \emph {et~al.}(2022)\citenamefont {Wu},
  \citenamefont {Liao},\ and\ \citenamefont {Foster}}]{Wu22}%
  \BibitemOpen
  \bibfield  {author} {\bibinfo {author} {\bibfnamefont {T.~C.}\ \bibnamefont
  {Wu}}, \bibinfo {author} {\bibfnamefont {Y.}~\bibnamefont {Liao}},\ and\
  \bibinfo {author} {\bibfnamefont {M.~S.}\ \bibnamefont {Foster}},\ }\bibfield
   {title} {\bibinfo {title} {Quantum interference of hydrodynamic modes in a
  dirty marginal fermi liquid},\ }\href
  {https://doi.org/10.1103/PhysRevB.106.155108} {\bibfield  {journal} {\bibinfo
   {journal} {Physical Review B}\ }\textbf {\bibinfo {volume} {106}},\ \bibinfo
  {pages} {155108} (\bibinfo {year} {2022})}\BibitemShut {NoStop}%
\bibitem [{\citenamefont {Jaggi}(1991)}]{Jaggi91}%
  \BibitemOpen
  \bibfield  {author} {\bibinfo {author} {\bibfnamefont {R.}~\bibnamefont
  {Jaggi}},\ }\bibfield  {title} {\bibinfo {title} {Electron-fluid model for
  the dc size effect},\ }\href {https://doi.org/10.1063/1.347315} {\bibfield
  {journal} {\bibinfo  {journal} {Journal of Applied Physics}\ }\textbf
  {\bibinfo {volume} {69}},\ \bibinfo {pages} {816} (\bibinfo {year}
  {1991})}\BibitemShut {NoStop}%
\bibitem [{\citenamefont {Shi}\ \emph {et~al.}(2014)\citenamefont {Shi},
  \citenamefont {Martin}, \citenamefont {Ebner}, \citenamefont {Zudov},
  \citenamefont {Pfeiffer},\ and\ \citenamefont {West}}]{Shi14}%
  \BibitemOpen
  \bibfield  {author} {\bibinfo {author} {\bibfnamefont {Q.}~\bibnamefont
  {Shi}}, \bibinfo {author} {\bibfnamefont {P.~D.}\ \bibnamefont {Martin}},
  \bibinfo {author} {\bibfnamefont {Q.~A.}\ \bibnamefont {Ebner}}, \bibinfo
  {author} {\bibfnamefont {M.~A.}\ \bibnamefont {Zudov}}, \bibinfo {author}
  {\bibfnamefont {L.~N.}\ \bibnamefont {Pfeiffer}},\ and\ \bibinfo {author}
  {\bibfnamefont {K.~W.}\ \bibnamefont {West}},\ }\bibfield  {title} {\bibinfo
  {title} {Colossal negative magnetoresistance in a two-dimensional electron
  gas},\ }\href {https://doi.org/10.1103/PhysRevB.89.201301} {\bibfield
  {journal} {\bibinfo  {journal} {Phys. Rev. B}\ }\textbf {\bibinfo {volume}
  {89}},\ \bibinfo {pages} {201301(R)} (\bibinfo {year} {2014})}\BibitemShut
  {NoStop}%
\bibitem [{\citenamefont {Renard}\ \emph {et~al.}(2008)\citenamefont {Renard},
  \citenamefont {Tkachenko}, \citenamefont {Tkachenko}, \citenamefont {Ota},
  \citenamefont {Kumada}, \citenamefont {Portal},\ and\ \citenamefont
  {Hirayama}}]{Renard08}%
  \BibitemOpen
  \bibfield  {author} {\bibinfo {author} {\bibfnamefont {V.~T.}\ \bibnamefont
  {Renard}}, \bibinfo {author} {\bibfnamefont {O.~A.}\ \bibnamefont
  {Tkachenko}}, \bibinfo {author} {\bibfnamefont {V.~A.}\ \bibnamefont
  {Tkachenko}}, \bibinfo {author} {\bibfnamefont {T.}~\bibnamefont {Ota}},
  \bibinfo {author} {\bibfnamefont {N.}~\bibnamefont {Kumada}}, \bibinfo
  {author} {\bibfnamefont {J.-C.}\ \bibnamefont {Portal}},\ and\ \bibinfo
  {author} {\bibfnamefont {Y.}~\bibnamefont {Hirayama}},\ }\bibfield  {title}
  {\bibinfo {title} {Boundary-mediated electron-electron interactions in
  quantum point contacts},\ }\href
  {https://doi.org/10.1103/PhysRevLett.100.186801} {\bibfield  {journal}
  {\bibinfo  {journal} {Physical Review Letters}\ }\textbf {\bibinfo {volume}
  {100}},\ \bibinfo {pages} {186801} (\bibinfo {year} {2008})}\BibitemShut
  {NoStop}%
\bibitem [{\citenamefont {Hatke}\ \emph {et~al.}(2012)\citenamefont {Hatke},
  \citenamefont {Zudov}, \citenamefont {Reno}, \citenamefont {Pfeiffer},\ and\
  \citenamefont {West}}]{Hatke12}%
  \BibitemOpen
  \bibfield  {author} {\bibinfo {author} {\bibfnamefont {A.~T.}\ \bibnamefont
  {Hatke}}, \bibinfo {author} {\bibfnamefont {M.~A.}\ \bibnamefont {Zudov}},
  \bibinfo {author} {\bibfnamefont {J.~L.}\ \bibnamefont {Reno}}, \bibinfo
  {author} {\bibfnamefont {L.~N.}\ \bibnamefont {Pfeiffer}},\ and\ \bibinfo
  {author} {\bibfnamefont {K.~W.}\ \bibnamefont {West}},\ }\bibfield  {title}
  {\bibinfo {title} {Giant negative magnetoresistance in high-mobility
  two-dimensional electron systems},\ }\href
  {https://doi.org/10.1103/PhysRevB.85.081304} {\bibfield  {journal} {\bibinfo
  {journal} {Physical Review B}\ }\textbf {\bibinfo {volume} {85}},\ \bibinfo
  {pages} {081304} (\bibinfo {year} {2012})}\BibitemShut {NoStop}%
\bibitem [{\citenamefont {Gusev}\ \emph {et~al.}(2018)\citenamefont {Gusev},
  \citenamefont {Levin}, \citenamefont {Levinson},\ and\ \citenamefont
  {Bakarov}}]{Gusev18}%
  \BibitemOpen
  \bibfield  {author} {\bibinfo {author} {\bibfnamefont {G.~M.}\ \bibnamefont
  {Gusev}}, \bibinfo {author} {\bibfnamefont {A.~D.}\ \bibnamefont {Levin}},
  \bibinfo {author} {\bibfnamefont {E.~V.}\ \bibnamefont {Levinson}},\ and\
  \bibinfo {author} {\bibfnamefont {A.~K.}\ \bibnamefont {Bakarov}},\
  }\bibfield  {title} {\bibinfo {title} {Viscous electron flow in mesoscopic
  two-dimensional electron gas},\ }\href {https://doi.org/10.1063/1.5020763}
  {\bibfield  {journal} {\bibinfo  {journal} {AIP Advances}\ }\textbf {\bibinfo
  {volume} {8}},\ \bibinfo {pages} {025318} (\bibinfo {year}
  {2018})}\BibitemShut {NoStop}%
\bibitem [{\citenamefont {Raichev}\ \emph {et~al.}(2020)\citenamefont
  {Raichev}, \citenamefont {Gusev}, \citenamefont {Levin},\ and\ \citenamefont
  {Bakarov}}]{Raichev20}%
  \BibitemOpen
  \bibfield  {author} {\bibinfo {author} {\bibfnamefont {O.~E.}\ \bibnamefont
  {Raichev}}, \bibinfo {author} {\bibfnamefont {G.~M.}\ \bibnamefont {Gusev}},
  \bibinfo {author} {\bibfnamefont {A.~D.}\ \bibnamefont {Levin}},\ and\
  \bibinfo {author} {\bibfnamefont {A.~K.}\ \bibnamefont {Bakarov}},\
  }\bibfield  {title} {\bibinfo {title} {Manifestations of classical size
  effect and electronic viscosity in the magnetoresistance of narrow
  two-dimensional conductors: Theory and experiment},\ }\href
  {https://doi.org/10.1103/PhysRevB.101.235314} {\bibfield  {journal} {\bibinfo
   {journal} {Phys. Rev. B}\ }\textbf {\bibinfo {volume} {101}},\ \bibinfo
  {pages} {235314} (\bibinfo {year} {2020})}\BibitemShut {NoStop}%
\bibitem [{\citenamefont {Hartnoll}\ \emph {et~al.}(2018)\citenamefont
  {Hartnoll}, \citenamefont {Lucas},\ and\ \citenamefont
  {Sachdev}}]{HartnollLucasSachdev18}%
  \BibitemOpen
  \bibfield  {author} {\bibinfo {author} {\bibfnamefont {S.~A.}\ \bibnamefont
  {Hartnoll}}, \bibinfo {author} {\bibfnamefont {A.}~\bibnamefont {Lucas}},\
  and\ \bibinfo {author} {\bibfnamefont {S.}~\bibnamefont {Sachdev}},\
  }\href@noop {} {\emph {\bibinfo {title} {Holographic Quantum Matter}}}\
  (\bibinfo  {publisher} {MIT Press},\ \bibinfo {address} {Cambridge, MA},\
  \bibinfo {year} {2018})\ \Eprint {https://arxiv.org/abs/1612.07324}
  {arXiv:1612.07324 [hep-th]} \BibitemShut {NoStop}%
\bibitem [{\citenamefont {Davison}\ \emph {et~al.}(2015)\citenamefont
  {Davison}, \citenamefont {Gout{\'e}raux},\ and\ \citenamefont
  {Hartnoll}}]{DavisonGouterauxHartnoll15}%
  \BibitemOpen
  \bibfield  {author} {\bibinfo {author} {\bibfnamefont {R.~A.}\ \bibnamefont
  {Davison}}, \bibinfo {author} {\bibfnamefont {B.}~\bibnamefont
  {Gout{\'e}raux}},\ and\ \bibinfo {author} {\bibfnamefont {S.~A.}\
  \bibnamefont {Hartnoll}},\ }\bibfield  {title} {\bibinfo {title} {Incoherent
  transport in clean quantum critical metals},\ }\href
  {https://doi.org/10.1007/JHEP10(2015)112} {\bibfield  {journal} {\bibinfo
  {journal} {Journal of High Energy Physics}\ }\textbf {\bibinfo {volume}
  {2015}},\ \bibinfo {pages} {112} (\bibinfo {year} {2015})},\ \Eprint
  {https://arxiv.org/abs/1507.07137} {arXiv:1507.07137 [hep-th]} \BibitemShut
  {NoStop}%
\bibitem [{\citenamefont {Damle}\ and\ \citenamefont
  {Sachdev}(1997)}]{DamleSachdev97}%
  \BibitemOpen
  \bibfield  {author} {\bibinfo {author} {\bibfnamefont {K.}~\bibnamefont
  {Damle}}\ and\ \bibinfo {author} {\bibfnamefont {S.}~\bibnamefont
  {Sachdev}},\ }\bibfield  {title} {\bibinfo {title} {Nonzero-temperature
  transport near quantum critical points},\ }\href
  {https://doi.org/10.1103/PhysRevB.56.8714} {\bibfield  {journal} {\bibinfo
  {journal} {Physical Review B}\ }\textbf {\bibinfo {volume} {56}},\ \bibinfo
  {pages} {8714} (\bibinfo {year} {1997})}\BibitemShut {NoStop}%
\end{thebibliography}
%

\end{document}